\documentclass[%
reprint,
superscriptaddress,
 amsmath,amssymb,
 aps,
floatfix,
]{revtex4-1}

\usepackage[dvipsnames,svgnames,x11names,hyperref]{xcolor}
\usepackage{siunitx}

\usepackage{graphicx}
\usepackage{dcolumn}
\usepackage{bm}
\usepackage{hyperref}
\usepackage[mathlines]{lineno}
\usepackage{mathrsfs}
\usepackage[normalem]{ulem}

\usepackage[margin=0.9in]{geometry}

\usepackage[normalem]{ulem}
\hypersetup{colorlinks=true, 
	linkcolor={blue!75!black!80!yellow},
	citecolor={blue!75!black!80!yellow}, 
	urlcolor=black 
	}
\makeatletter \renewcommand\@make@capt@title[2]{%
\@ifx@empty\float@link{\@firstofone}{\expandafter\href\expandafter{\float@link}}%
\sffamily{\textbf{#1}}\@caption@fignum@sep#2 }

\begin{document}
\preprint{APS/123-QED}

\title{Selective acoustic control of photon-mediated qubit-qubit interactions}

\author{Tom\'{a}\v{s} Neuman}
\email{tomasneuman@seas.harvard.edu}
\affiliation{Harvard John A. Paulson School of Engineering and Applied Sciences, Harvard University, Cambridge, MA 02138, USA}
\author{Matthew Trusheim}
\affiliation{Harvard John A. Paulson School of Engineering and Applied Sciences, Harvard University, Cambridge, MA 02138, USA}
\email{mtrusheim@fas.harvard.edu}
\author{Prineha Narang}
\email{prineha@seas.harvard.edu}
\affiliation{Harvard John A. Paulson School of Engineering and Applied Sciences, Harvard University, Cambridge, MA 02138, USA}

\begin{abstract}
Quantum technologies such as quantum sensing, quantum imaging, quantum communications, and quantum computing rely on the ability to actively manipulate the quantum state of light and matter. Quantum emitters, such as color centers trapped in solids, are a useful platform for the realization of elementary building blocks (qubits) of quantum information systems. In particular, the modular nature of such solid-state devices opens up the possibility to connect them into quantum networks and create non-classical states of light shared among many qubits. The function of a quantum network relies on efficient and controllable interactions among individual qubits. In this context, we present a scheme where optically active qubits of differing excitation energies are mutually coupled via a dispersive interaction with a shared mode of an optical cavity. This generally off-resonant interaction is prohibitive of direct exchange of information among the qubits. However, we propose a scheme in which by acoustically modulating the qubit excitation energies it is in fact possible to tune to resonance a pre-selected pair of qubits and thus open a communication channel between them. This method potentially enables fast ($\sim$ns) and parallelizable on-demand control of a large number of physical qubits. We develop an analytical and a numerical theoretical model demonstrating this principle and suggest feasible experimental scenarios to test the theoretical predictions.
\end{abstract}
\date{\today}

\maketitle

\section{Introduction}

\begin{figure}[t]
    \centering
    \includegraphics[scale=0.6]{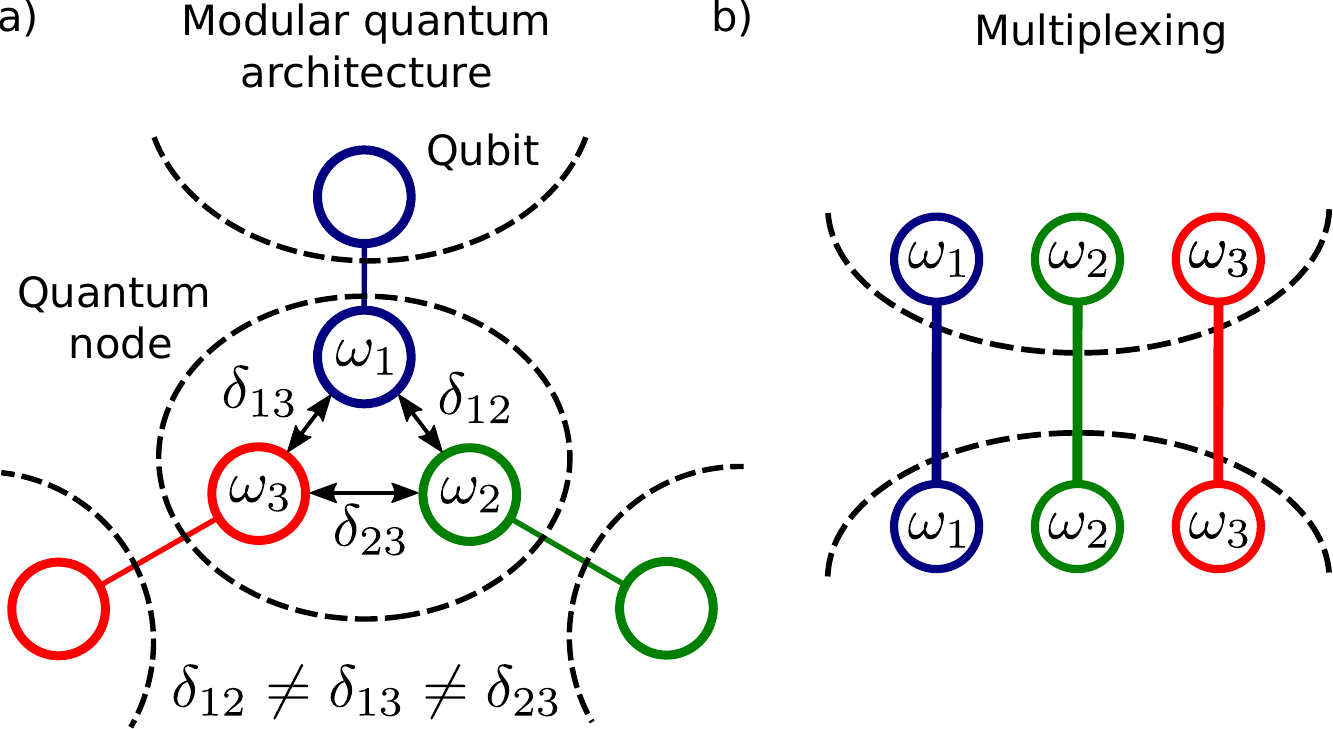}
    \caption{Schematically shown are (a) a modular quantum architecture and (b) quantum multiplexing. (a) A quantum node composed of a set of qubits of potentially different excitation frequencies $\omega_1$, $\omega_2$ and $\omega_3$ (with $\delta_{ij}=\omega_i-\omega_j$) is embedded in a quantum network. The method described in this work can, for example, selectively trigger interactions between selected qubit pairs. (b) Using qubits of different optical frequencies enables spectral multiplexing of an optically active quantum node.}
    \label{fig:figINT}
\end{figure}

Control over the quantum states of light and matter is a central component of many applications, ranging from quantum computation and communication to energy transfer and the realization of many-body phenomena. A relevant and timely application in which single photons interact with single quanta of matter is the development of quantum information processing architectures based on optically-active spin systems \cite{Awschalom2018-en}. Quantum emitters coupled to optical modes are key components in quantum networks where entanglement is distributed across distances \cite{Kimble2008-cu,Wehner2018-wx}, and in the development of interconnected, modular quantum computers \cite{Jiang2007-hr,Hucul2015-td}, as schematically depicted in Fig.\,\ref{fig:figINT}(a). Entanglement in these systems is formed through two general interactions: a long-distance optical interaction, and a local interaction that joins optically-entangled qubits together to form a large, potentially useful quantum state. Many local interactions exist for different physical qubits, including magnetic dipolar coupling of spins \cite{Dutt2007-tj,Robledo2011-fw,Dolde2013-vh}, Coulombic interaction between ions \cite{Turchette1998-kf,Monroe2013-mq}, or photon-mediated interactions between atomic(-like) emitters \cite{Welte2018-us, Evans2018-vh}. 
Qubits in solid-state systems uniquely offer the potential for interaction via the surrounding crystalline elastic deformations (acoustic phonons) than can be confined into phononic cavities \cite{gavartin2011optomechcouple,Aspelmeyer2012-sy,Galland2014-cl,Safavi-Naeini2014-af, fang2016optical, golter2016optomechNV,burek2016diamondoptomech, lemonde2018phononnetworks, moores2018quantumacoustic, whiteley2019spin, maccabe2019phononic}. These phonon-mediated interactions require qubit susceptibility to local strain, which has been demonstrated in a variety of solid-state quantum-emitter systems \cite{ovartchaiyapong2014dynamic,schuetz2015universaltransducers,macquarrie2015contdecoupling,Jahnke2015elphonon, lemonde2018phononnetworks, chen2019engelphon, li2019honeycomb, chen2019engelphon, maity2019coherent}. At the same time, strain susceptibility induces local variation in qubit energy levels \cite{Lindner2018-ka}, leading to an inhomogeneous broadening that can prevent efficient, resonant interactions.

Here we describe a scheme that takes advantage of this inhomogeneous broadening and exploits the uniquely strong mechanical response of solid-state emitters to quickly and selectively tune optical interactions between multiple qubits. Importantly, we show that the application of a high-frequency strain field can enable the efficient coherent interaction of spectrally mismatched emitters within a cavity mode, potentially enabling a two-qubit gate even in the presence of disorder. Finally, we show that the mechanical interaction can be used to engineer pairwise coupling between emitters within a larger cluster with high speed, enabling for example spectral multiplexing of an optically-active quantum node [Fig.\,\ref{fig:figINT}(b)].


The approach presented here addresses key issues in the development of quantum networks. The ability to engineer interactions between spectrally distinct emitters relaxes the requirements on homogeneity between quantum nodes. Photon-mediated entanglement-at-a-distance generally requires a Bell measurement between indistinguishable photons to perform an entanglement swap \cite{Simon2003-zu,Moehring2007-qw,Bernien2013-qt}, although high detector resolution can alleviate this requirement to an extent \cite{Dyckovsky2012-tf}. If each node has a single optically-active qubit, all nodes must operate within the allowed spectral band, placing limits on the distribution of usable quantum emitters. By enabling local entanglement between spectrally distinct emitters, usable bandwidth is increased and wavelength-domain multiplexing becomes allowed. This has the potential to increase entanglement generation rate significantly within an appropriate architecture \cite{Lo_Piparo2019-sa}. Finally, the ability to connect locally to additional emitters enables an increase in the number of qubits per node.  

A large variety of solid-state quantum emitters have been studied, with a range of advantages and drawbacks for each system \cite{Atature2018-oh}. Central to the proposed scheme are two elements: strong interaction between the emitter optical transition and a cavity mode, and strong interaction between the optical transition energies and mechanical strain (phonons). Recent work with quantum emitters in diamond \cite{Evans2018-vh,Nguyen2019-yd} has demonstrated coupling to optical cavities with cooperativity $C$ $>$ 100, enabling the high-efficiency light-matter interaction necessary for quantum networking applications. Measurements of the strain interaction of these systems have also revealed a high susceptibility of 1 PHz/strain \cite{Meesala2018-zl}. We therefore consider solid-state systems with similar optical and strain properties in our analysis, though the results are extensible across the wide range of solid-state quantum systems. 

This Article is structured as follows. In Section\,\ref{sec:model} we present a theoretical model of the acoustically induced resonant photon-mediated interactions. We develop both a numerical model in Subsection\,\ref{subsec:nummod} and an analytical model in Subsection\,\ref{subsec:anmod} that elucidates the physical mechanism of qubit coupling and allows for choosing optimal parameters of the acoustic drive. We then demonstrate acoustically-driven interaction between spectrally detuned qubits by numerically solving this model, and extending it to the case of selective pairwise interactions within a multi-qubit cluster in Section\,\ref{sec:demonstration}.

\section{Model}\label{sec:model}
\subsection{Full model}\label{subsec:nummod}
\begin{figure}[h]
    \centering
    \includegraphics[scale=0.5]{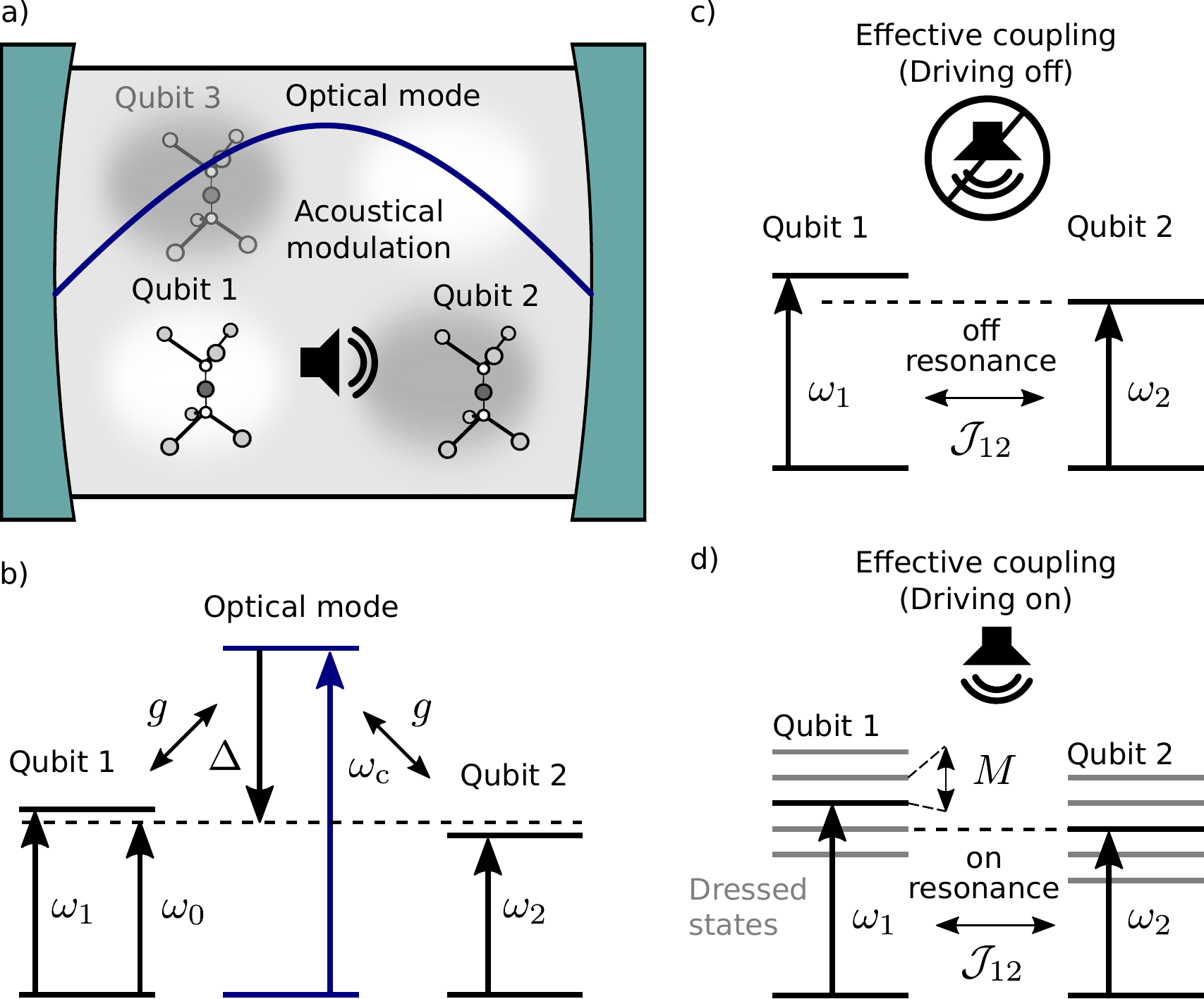}
    \caption{The principle of acoustically induced qubit-qubit interaction. (a) A set of two (or three) optically active qubits (such as diamond impurities) is placed into a photonic cavity. All of the emitters interact with an optical mode of the cavity. The excitation energies of the emitters can be modulated by acoustically driving the qubit hosting medium. (b) Schematic diagram of energy levels involved in the model of two qubits coupled to a common cavity mode. Both qubits are described as two-level systems of respective excitation frequencies $\omega_1$ and $\omega_2$ that are not equal and both are detuned from the cavity resonance frequency $\omega_{\rm c}$ by $\approx \Delta$. The qubit-cavity coupling is characterised by a coupling rate $g$ equal for both qubits. (c) Upon effective elimination of the cavity mode the two qubits are non-resonantly coupled via an effective rate $\mathcal{J}_{12}$ if the acoustic drive is turned off. (d) The acoustic drive of frequency $M$ induces phonon side bands (dressed states) around the bare qubit frequencies which can allow for sideband-mediated resonant qubit-qubit interaction.}
    \label{fig:fig1}
\end{figure}
The model system consists of several optically active qubits (two-level systems) embedded in an optical cavity. These qubits can be, for example, point defects in solids such as the (negatively charged) silicon-vacancy (SiV$^-$) color centers in diamond, while the cavity can formed by a number of implementations e.g. photonic-crystal structuring of the surrounding crystal lattice. We schematically show such a system in Fig.\,\ref{fig:fig1}(a). The optical transition frequencies $\omega_{i}=\omega_0+\delta_i$ (with $i=1,\,2,\,3$) of the individual qubits are similar, but not equal to each other ($\delta_i\ll \omega_0$, $\delta_i\neq \delta_j$ for $i\neq j$), and are detuned from the frequency $\omega_{\rm c}$ of a single dominant mode of the cavity so that $\Delta\gg \delta_i$ with $\Delta=\omega_0-\omega_{\rm c}$. The respective qubit-qubit detunings $\delta_{i}$ are determined by the inhomogeneous broadening that is naturally present in the system due to local strain or other imperfections within the surrounding crystal, or can be artificially induced, for example by locally applying strain or electromagnetic fields to the qubits. The Hamiltonian $H_{0}$ of this model system can be expressed as:
\begin{align}
    H_{0}=\sum_i \hbar \omega_i \sigma_i^\dagger \sigma_i + \hbar \omega_{\rm c} a^\dagger a.\label{eq:h0}
\end{align}
Here $\sigma_i$ ($\sigma_i^\dagger$) is the lowering (raising) operator of the qubit $i$, and $a$ ($a^\dagger$) is the annihilation (creation) operator representing the cavity mode. The qubits are coupled to the cavity mode via their dipole-allowed transitions. We describe this interaction in the rotating-wave approximation via the Jaynes-Cummings coupling term: 
\begin{align}
    H_{\rm J-C}=\hbar g\sum_{i} (\sigma_i a^\dagger + \sigma_i^\dagger a),\label{eq:hjc}
\end{align}
where $g_{i}=g$ are the respective coupling rates that we assume identical for simplicity. The qubit-cavity interaction in $H_{\rm J-C}$ together with $H_0$ ensures that the qubit-cavity interaction is in the dispersive regime that will allow us to introduce effective qubit-qubit interactions by eliminating the cavity, as described in Section\,\ref{sec:effectivestatic}. The level structure of the system described is outlined in Fig.\,\ref{fig:fig1}(b). 

To control the effective qubit-qubit interactions we apply a coherent acoustic drive of frequency $M$ which produces the respective modulated qubit excitation energies $\tilde{\omega}_i(t)$ \cite{noel1998freqmod, ashhab2007drivetheory, beaudoin2012sidebands, strand2013sideband, silveri2017quantum} according to the qubit strain susceptibilities \cite{Oliver1653machzender, golter2016optomechNV, maity2018alignment, udvarheliy2018spinstrain, Meesala2018-zl, chen2018orbital, whiteley2019spin, dmitriev2019spinresonance} as:
\begin{align}
    \tilde \omega_i (t)=\omega_{i}+\mathcal{D}_i \cos(Mt+\phi_i),\label{eq:ommod}
\end{align}
where $\mathcal{D}_i$ is the drive amplitude and $\phi_i$ is the phase offset of the drive. This drive can be induced in the crystal by e.g. contact with a piezo-electric transducer or capacitive drive, and is externally controllable in amplitude and frequency. 
The driven system is described by the Hamiltonian $\tilde{H}(t)=\tilde H_0(t) + H_{\rm J-C}$ with:
\begin{align}
    \tilde H_0(t)=\sum_i \hbar \tilde\omega_i(t) \sigma_i^\dagger \sigma_i + \hbar \omega_{\rm c} a^\dagger a.\label{eq:h0t}
\end{align}

We assume that the qubits and the cavity mode experience both population decay and pure dephasing processes due to the environment. We describe the dynamics of such a lossy system in the framework of the Lindblad master equation for the system's density matrix $\rho$:
\begin{align}
    \frac{{\rm d}\rho}{{\rm d}t}=\frac{1}{{\rm i}\hbar}[ \tilde H(t), \rho] + \sum_{i }\gamma_{c_i}\mathcal{L}_{c_i}(\rho).\label{eq:master}
\end{align}
Here 
\begin{align}
    \gamma_{c_i}\mathcal{L}_{c_i}(\rho)=\frac{\gamma_{c_i}}{2} \left(2 c_i\rho c_i^\dagger-\lbrace c_i^\dagger c_i,\rho \rbrace \right), 
\end{align}
with $c_i\in \{ a,\sigma_{j}, \sigma_{j}^\dagger\sigma_j \}$. For $c_i\in \{ a,\sigma_{j}\}$, $\gamma_{c_i}$ are the decay rates of the respective excitations, and for $c_i\in \{ \sigma_{j}^\dagger\sigma_j \}$, $\gamma_{c_i}$ represent the pure dephasing rates. We choose the respective rates so that $\Delta > \gamma_a, \gamma_{\sigma_i}$ and ensure that the cavity does not induce strong decay of the emitters due to the Purcell effect. In the following sections, we solve the master equation Eq.\,\eqref{eq:master} numerically using standard methods \cite{Breuer2003}. Below we analyze the properties of the full numerical model and develop an effective analytical model describing the mechanism of the acoustically induced photon-mediated qubit-qubit interaction. 
\subsection{Analytical model}\label{subsec:anmod}
To elucidate the physical mechanism of the photon-mediated coupling, we now briefly discuss the origin of the cavity-mediated qubit-qubit interaction. After that we detail how the acoustic drive can be exploited to dynamically tune a resonance leading to an efficient coupling between a chosen pair of qubits.

\subsubsection{Effective qubit-qubit coupling in the dispersive regime}\label{sec:effectivestatic}

In the dispersive regime the optical mode of the cavity effectively mediates the qubit-qubit interaction which can be shown by applying to $H=H_0+H_{\rm J-C}$ [Eq.\,\eqref{eq:h0} and Eq.\,\eqref{eq:hjc}] the unitary transformation represented by the operator $U$ \cite{zheng2000twoatentangl, blais2004dispersive, majer2007cavitybus},
\begin{align}
U=\exp \left( \sum_i\frac{g}{\Delta_i}\left [a^\dagger \sigma_i- a\sigma_i^\dagger\right]\right),
\end{align}
where $\Delta_i=\Delta+\delta_i$. This can be done using the Hausdorff identity
\begin{align}
e^{-X}H e^{ X}=H+ [H,X]+\frac{1}{2}[[H, X], X]+\ldots ,   
\end{align}
where $X$ is an arbitrary operator proportional to a small parameter $g/\Delta$.
The transformed Hamiltonian $ \hat H$ is (retaining only terms to the first order in $g/\Delta$)
 \begin{align}
     \hat H&=U^\dagger H U\nonumber\\
     &\approx \sum_{i}\hbar \omega_i \sigma^\dagger_i\sigma_i+\hbar\omega_{\rm c}a^\dagger a\nonumber\\
     &+\sum_{i,j}\hbar \mathcal{J}_{ij}  \sigma^\dagger_i\sigma_{j}  + \sum_j\hbar \mathcal{J}_{jj}a^\dagger a (2\sigma^\dagger_j \sigma_j-I)\nonumber\\
     &\approx \sum_{i}\hbar \omega_i \sigma^\dagger_i\sigma_i+\hbar\omega_{\rm c}a^\dagger a+\sum_{i,j}\hbar \mathcal{J}_{ij}  \sigma^\dagger_i\sigma_{j}.\label{eq:hath}
 \end{align}
where $\mathcal{J}_{ij}\equiv \frac{g^2(\Delta_i+\Delta_j)}{2\Delta_{i}\Delta_j} \approx \frac{g^2}{\Delta}$, and in the last step we have assumed that in the dispersive regime the cavity is only weakly populated ($a^\dagger a \approx 0$) and that $\frac{g^2}{\Delta}\ll \omega_i, \omega_{\rm c}, |\Delta_i|$. We see that in the last line of Eq.\,\eqref{eq:hath} the cavity dynamics is decoupled from the qubit dynamics which allows us to define the effective qubit-only Hamiltonian:
\begin{align}
    H_{\rm eff}=\sum_{i}\hbar \omega_i \sigma^\dagger_i\sigma_i+ \sum_{ij}\hbar\mathcal{J}_{ij}  \sigma^\dagger_i\sigma_{j}.\label{eq:Heff}
\end{align}
Equation\,\eqref{eq:Heff} describes a direct qubit-qubit interaction which can lead to hybridization of qubit excited states provided that the qubit-qubit coupling is stronger than the detuning of the respective qubit frequencies $\mathcal{J}_{ij}\approx \mathcal{J}>|\omega_i-\omega_j|$ (with $\mathcal{J}=g^2/\Delta$ and $i\neq j$). However, if $\mathcal{J}<|\omega_i-\omega_j|$ and $\gamma_{\sigma_i}, \gamma_{\sigma^\dagger_i\sigma_i}<|\omega_i-\omega_j|$, as in our model scenario of spectrally inhomogeneous emitters, the two qubits interact non-resonantly and are effectively isolated as schematically shown in Fig.\,\ref{fig:fig1}(c) for a pair of interacting qubits. The qubits are thus effectively non-interacting in the absence of the acoustic drive. 

\subsubsection{Effect of the acoustic drive}\label{subsubsec:adrive}
\begin{figure}
    \centering
    \includegraphics[scale=0.77]{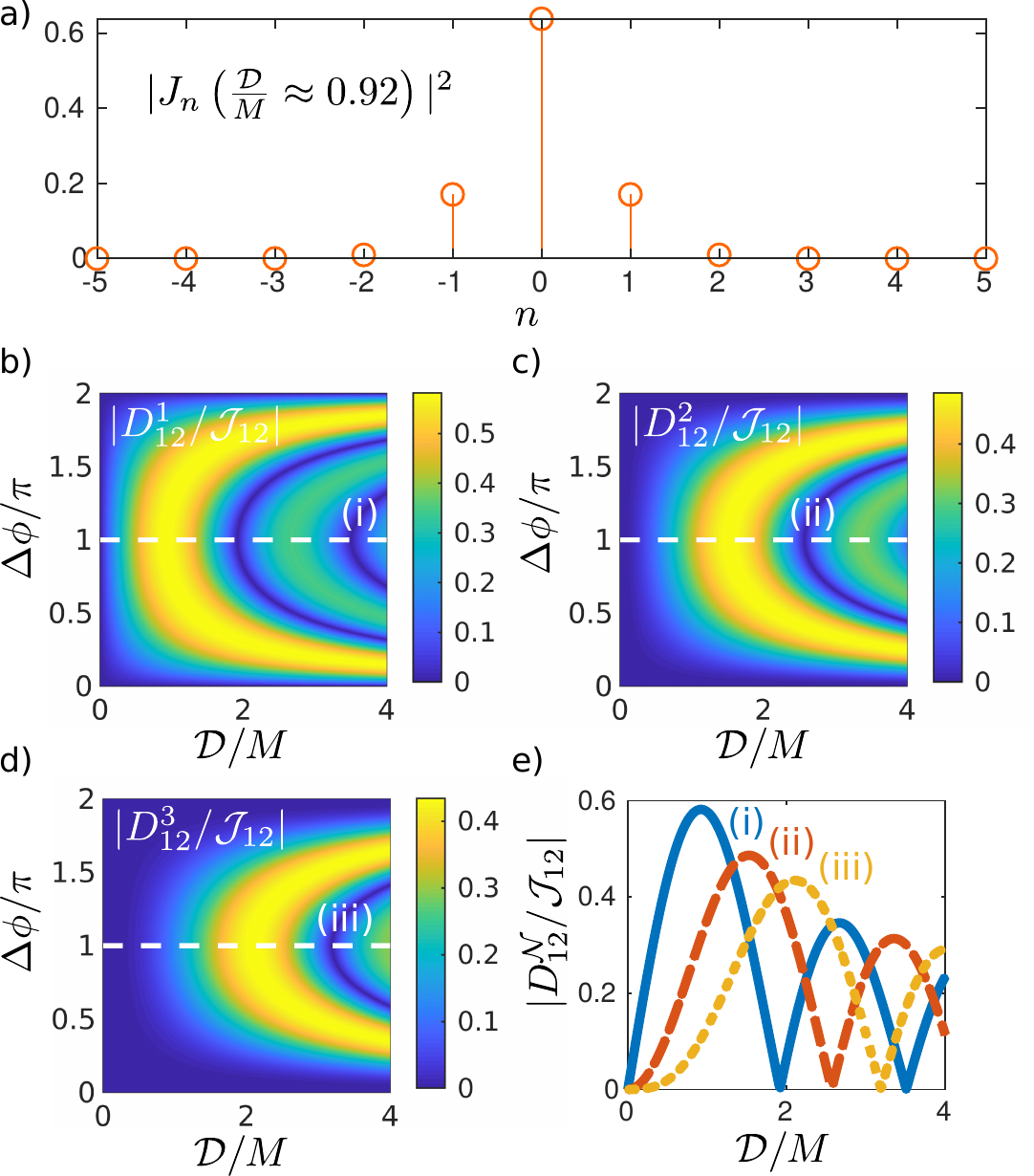}
    \caption{Analytical results characterising the mechanism of the acoustically induced resonant qubit-qubit interaction. (a) Relative amplitudes of phonon sidebands of a frequency modulated qubit. (b-e) Effective qubit-qubit coupling $D_{12}^{\mathcal{N}}$ (normalized to $\mathcal{J}_{12}$) as a function of $\mathcal{D}/M$ (assuming $\mathcal{D}_1=\mathcal{D}_2=\mathcal{D}$) and relative phase between the drive of qubit 1 and qubit 2 ($\Delta\phi=\phi_1-\phi_2$) when $\mathcal{N}\times M=\delta_{12}$ for (b) $\mathcal{N}=1$, (c) $\mathcal{N}=2$, and (d) $\mathcal{N}=3$. In (e) we show the cuts along the white-dashed lines in (a-c) labelled respectively as (i), (ii), and (iii). }
    \label{fig:fig0}
\end{figure}
As we schematically show in Fig.\,\ref{fig:fig1}(d) the main effect of the acoustic drive is to introduce dressed states that appear as sidebands around the original qubit excitation frequency $\omega_i$, equidistantly positioned at frequencies $\omega_{ i}+n M$ (with $n$ an integer) \cite{Oliver1653machzender, ashhab2007drivetheory, chen2018orbital, whiteley2019spin, dmitriev2019spinresonance}. 
These sidebands can be experimentally identified in the emission spectra of acoustically driven qubits \cite{chen2018orbital} in the form of a frequency comb formed by a number of equidistant peaks (see Appendix\,\ref{app:emspec} for details). The amplitude of a sideband $n$ of the spectrum is determined by $\propto |J_n(\mathcal{D}/M)|^2$ where $J_n(x)$ is the Bessel function of the first kind \cite{abramowitz65functions}. As an example, we show the frequency comb obtained for a single qubit modulated at an amplitude of $\mathcal{D}/M\approx 0.92$ in Fig.\,\ref{fig:fig0}(a) (we neglect the spectral line width). Notice that the higher-order sidebands are not significant for this modulation amplitude, limiting the total bandwidth of the qubit transitions.

By choosing the driving frequency $M$ such that $M=|\omega_i-\omega_j|$ it is thus possible to tune to resonance the acoustic side-bands of two distinct qubits and thus turn on a resonant qubit-qubit interaction. To elucidate this mechanism, we consider the following time-dependent effective Hamiltonian $\tilde H_{\rm eff}(t)$:
\begin{align}
    \tilde H_{\rm eff}(t)=\sum_i \hbar \tilde\omega_{i}(t) \sigma_i ^\dagger\sigma_{i}+\sum_{ij} \hbar\mathcal{J}_{ij} \sigma_i^\dagger \sigma_j,\label{eq:hefft}
\end{align}
where the qubit excitation energies are modulated according to Eq.\,\eqref{eq:ommod} and we have neglected any time dependence of the qubit-qubit coupling [$\mathcal{J}_{ij}(t)\approx \mathcal{J}_{ij}$]. The Hamiltonian in Eq.\,\eqref{eq:hefft} can be transformed into an interaction picture $\hat H_{\rm eff}(t)=U_t^\dagger \tilde H_{\rm eff}(t) U_t$ in which the time-dependence appears explicitly in the interaction term and the qubit excitation energies become time-independent. This can be accomplished by choosing 
\begin{align}
    U_{t}&=\exp \left( -{\rm i}\int_0^t\sum_j [\tilde\omega_{ j}(\tau)+\mathcal{J}_{jj}]\sigma_{j}^\dagger \sigma_{ j}  {\rm d}\tau \right)\nonumber\\
    &=\exp \left( -{\rm i}\sum_j\left[(\omega_{j}+\mathcal{J}_{jj})t +\frac{\mathcal{D}_j}{M}\sin(M t+\phi_{j})\right]\sigma_{j}^\dagger \sigma_{j} \right).
\end{align}
After applying $U_t$, the interaction-picture qubit lowering operator $\tilde \sigma_j$ becomes time-dependent:
\begin{align}
    \tilde\sigma_{j}(t)=\sigma_j e^{-{\rm i}\left[(\omega_j+\mathcal{J}_{jj})t+\frac{\mathcal{D}_j}{M}\sin(M t+\phi_j)\right]}\nonumber\\
    =\sigma_j e^{-{\rm i}(\omega_j+\mathcal{J}_{jj})t}\sum_n J_{n}\left(\frac{\mathcal{D}_j}{M}\right)e^{-{\rm i}n (M t + \phi_j)}.\label{eq:intpicture}
\end{align}
The transformed Hamiltonian becomes
\begin{align}
    \hat H_{\rm eff}(t)&=\sum_{i\neq j}\hbar\mathcal{J}_{ij}\sigma_{i}^\dagger \sigma_j e^{-{\rm i}\delta_{ij}t}\sum_{m n} f^{ij}_{mn}(t),
\end{align}
where we have defined $\delta_{ij}=\delta_j-\delta_i +\mathcal{J}_{jj}-\mathcal{J}_{ii}$ and the time-dependent factor $f_{mn}(t)$:
\begin{align}
    f^{ij}_{mn}(t)=J_m\left( \frac{\mathcal{D}_i}{M} \right)J_n\left( \frac{\mathcal{D}_j}{M} \right)e^{{\rm i}(m-n)Mt}e^{{\rm i}(m\phi_i-n\phi_j)}.
\end{align}
The function $f_{mn}(t)$ can be simplified in the secular approximation if we assume that the modulation frequency $M$ is resonant with the difference $\delta_{ij}$ of the transition frequencies of a given pair of qubits (further denoted as qubit $i=1$ and qubit $j=2$). We then retain only the time-independent terms that most dominantly contribute to the dynamics of the system provided that $|J_{\rm 12}|<|\delta_{12}|\approx \mathcal{N}\times M$ (with $\mathcal{N}$ an integer). In this approximation, the effective coupling between the pair of qubits $G^{\mathcal{N}}_{12}$ is given by
\begin{align}
    G^{\mathcal{N}}_{12}&=\mathcal{J}_{12}\sum_n e^{-{\rm i}\delta_{12}t}f^{12}_{(n+\mathcal{N})n}(t)\nonumber\\
    &=\mathcal{J}_{12}e^{{\rm i}\mathcal{N}\phi_1}\sum_n J_{n+\mathcal{N}}\left( \frac{\mathcal{D}_1}{M} \right)J_n\left( \frac{\mathcal{D}_2}{M} \right)e^{{\rm i}n\Delta\phi}\nonumber\\
    &=\mathcal{J}_{12} e^{{\rm i}\mathcal{N}\phi_1} e^{{\rm i}\mathcal{N}\psi} J_{\mathcal{N}}(z).\label{eq:g12}
\end{align}
with $\Delta\phi=\phi_1-\phi_2$,
\begin{align}
    z=\sqrt{\left(\mathcal{D}_1/M\right)^2+\left(\mathcal{D}_2/M\right)^2-2\mathcal{D}_1\mathcal{D}_2/M^2\cos(\Delta\phi)},
\end{align} 
and 
\begin{align}
    \sin(\psi)=\mathcal{D}_2/(M z)\sin(\Delta\phi)\quad (0<\Delta\phi \le \pi),
\end{align} 
which follows from the Graf's addition theorem (see Eq. 9.1.79 of \cite{abramowitz65functions} for more details).
We finally approximate the effective Hamiltonian describing the interaction between the two qubits by
\begin{align}
     H_{\rm eff, 2}=\hbar G^\mathcal{N}_{12} \sigma^\dagger_1\sigma_2+{\rm H.c.}\label{eq:efham}
\end{align}
Here, the qubit pair does not interact with the remaining off-resonant qubits ($G_{13}^{\mathcal{N}}\approx G_{23}^{\mathcal{N}}\approx 0$ in this approximation), while the interaction between the selected qubits is resonant and thus allows for an efficient qubit-qubit coupling. Furthermore, by tuning the drive frequency $M$ it is possible to selectively tune to resonance distinct qubit pairs. All qubit pairs that share the same detuning, or small-integer multiples of that detuning, will be efficiently coupled in the presence of the acoustic drive. For selective interactions, it is therefore important to ensure that the differences of excitation frequencies of qubit pairs are not each other's integer multiples ($|\delta_{ij}|\neq m |\delta_{kl}|$ for $\{ij\}\neq\{kl\}$), or this effect can be employed to generate parallel interactions between multiple qubits. 

This requirement can be relaxed for example, if one pair is indeed an integer multiple, but is a high integer multiple and the drive is not too strong. Such a situation arises naturally for the optimal choice of the pumping amplitude $\mathcal{D}/M\approx 0.92$ as in this case higher-order sidebands of the qubit resonance are weak as shown in Fig.\,\ref{fig:fig0}(a) (the amplitude of the $n=2$ peak reaches about 6\% of the $n=1$ peak amplitude). Another way to suppress the acoustically mediated tuning is to ensure that qubits that are supposed to be isolated are driven with equal phase. 

Finally, the secular approximation allowing for the selective interaction of qubit pairs is expected to break down if the photon-mediated qubit-qubit interaction $\mathcal{J}$ is comparable to $|\delta_{ij}-\delta_{kl}|$. In this case the resonant tuning of qubit frequencies is not necessary to trigger the qubit-qubit energy transfer as described in Subsection\,\ref{sec:effectivestatic}. We provide more details about the breakdown of the secular approximation in Appendix\,\ref{app:secular}.
\begin{figure}[h!]
    \centering
    \includegraphics[scale=0.7]{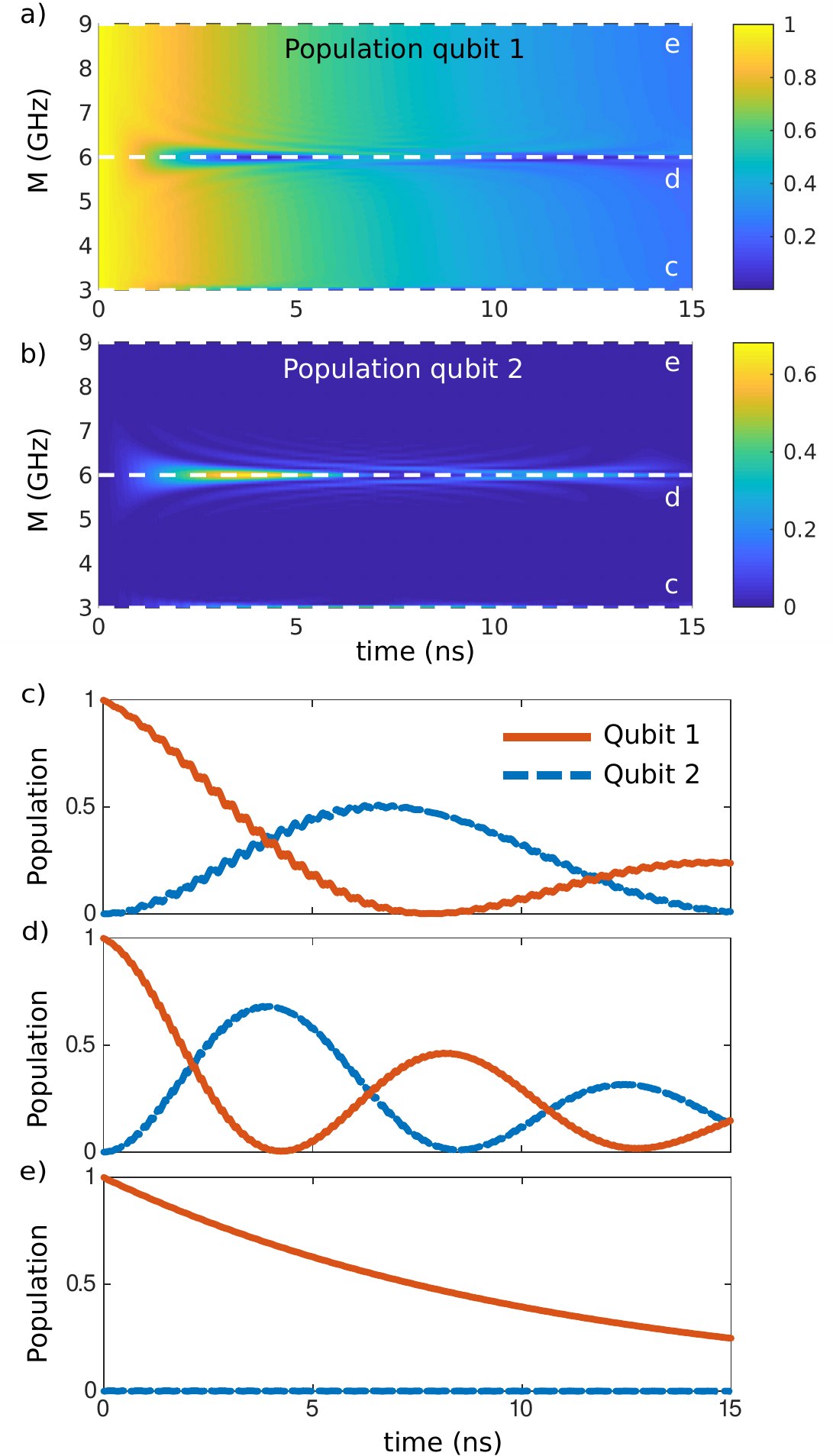}
    \caption{Time evolution of populations of two distinct quantum emitters (qubits) interacting dispersively with a photonic cavity. (a,b) Populations of the respective emitters as a function of time and frequency of the pumping acoustic wave. We assume that the two emitters are modulated with opposite phase ($\phi_1=\phi_2+\pi$) and that the amplitude of the modulation is equal for both emitters $\mathcal{D}_1/M=\mathcal{D}_2/M\approx 0.92$, which maximizes the value of $J_{1}(2\mathcal{D}_{1(2)}/M)$ and hence leads to an efficient qubit-qubit coupling. In (c,d,e) we plot cuts along the dashed lines shown in (a,b) and show how the populations (orange line for qubit 1 and blue dashed line for qubit 2) evolve if (a) $2M=\delta_{12}$, (b) $M=\delta_{12}$, and (c) $M=\frac{3}{2}\delta_{12}$. If an integer multiple of the modulation frequency $M$ is tuned to $\delta_{12}$, the two emitters are resonantly coupled and an exchange of the emitter's population occurs (the system is in the \textit{on} state) which can be observed as a Rabi oscillation in (c,d). If $M$ is not compatible with $\delta_{12}$, no exchange of populations occurs (the system is in the \textit{off} state) as shown in (e). }
    \label{fig:fig2}
\end{figure}

To show how the magnitude of the qubit-qubit interaction can be optimized, we plot the absolute value of $G^{\mathcal{N}}_{12}$ normalized to $\mathcal{J}_{\rm 12}$ in Fig.\,\ref{fig:fig0}(b-d) for (b) $\mathcal{N}=1$, (c) $\mathcal{N}=2$, and (d) $\mathcal{N}=3$ as a function of $\mathcal{D}_1=\mathcal{D}_2=\mathcal{D}$ normalized to the drive frequency $M$. We see that if the frequency modulation is in-phase, $\Delta\phi=0$, the two qubits decouple inasmuch as relative modulation of the qubit frequencies is necessary to trigger the qubit-qubit interaction. For $\Delta\phi=\pi$ the out-of-phase modulation of the two qubit excitation frequencies maximizes the effect of the mechanical drive and leads to the most efficient qubit-qubit interaction. In Fig.\,\ref{fig:fig0}(e) we plot the line cuts marked in Fig.\,\ref{fig:fig0}(b-d) [that can be analytically expressed as $|J_{\mathcal{N}}(2\mathcal{D}/M)|$ for (i) $\mathcal{N}=1$, (ii) $\mathcal{N}=2$, and (iii) $\mathcal{N}=3$, respectively] to show that the qubit-qubit interaction is most efficient for $\mathcal{N}=1$ and occurs when $\mathcal{D}/M\approx 0.92$. We take advantage of this result in Section\,\ref{sec:demonstration} where we set the parameters of the acoustic drive to optimize the coupling. In Appendix\,\ref{app:physparam} we discuss the physical conditions allowing for the optimal acoustic drive for the case of diamond emitters.

In the following section we discuss numerical results based on parameters achievable in diamond-based devices demonstrating the mechanism of acoustically induced selective resonant qubit-qubit interactions.

\section{Numerical solution}\label{sec:demonstration}
To demonstrate the principle of acoustically induced resonant qubit-qubit interaction we numerically solve the model outlined in Section\,\ref{sec:model} [Eq.\,\eqref{eq:hjc}, Eq.\,\eqref{eq:h0t}, and Eq.\,\eqref{eq:master}]. We choose the model parameters so that they represent a realistic system of qubits in the form of point defects in diamond (such as a SiV$^-$). \cite{Evans2018-vh}. In particular we consider a system consisting of two or three qubits of their respective frequencies $\delta_1=0$\,GHz, $\delta_2=2\pi\times 6$\,GHz and $\delta_{3}=2\pi\times 8$\,GHz. The cavity is detuned from the qubits such that $\Delta=2\pi\times 250$\,GHz. We assume that all the qubits are coupled to the cavity with an equal coupling strength of $g=2\pi \times 5$\,GHz. We further assume that the cavity resonance is broadened by $\gamma_a=2\pi\times 25$\,GHz and the intrinsic qubit decay rates are assumed to be identical and equal to $\gamma_{\sigma_i}=\gamma_{\sigma}=2\pi\times 5$\,MHz. We set the pure dephasing $\gamma_{\sigma^\dagger_i\sigma_i}=\gamma_{\sigma^\dagger\sigma}=0$\,Hz. With this choice of parameters we are thus operating in the bad-cavity limit. We have chosen qubit decay rate that overestimates realistic lifetimes of excited states of optical emitters but we use it here to demonstrate the principle of the scheme proposed. We discuss the effect of more realistic choice of the intrinsic decay and dephasing in Appendix\,\ref{app:param}. Nevertheless, we stress that the mechanism proposed here is general and does not only apply to diamond impurities, but can be generalized to any system featuring a similar coupling scheme. 

The dynamics that we obtain numerically is shown in Fig.\,\ref{fig:fig2}. We assume that qubit 1 is initially in the excited state $|{\rm e}\rangle$, and that qubit 2 and the cavity mode are in their respective ground states, $|{\rm g}\rangle $ and $|0\rangle$. The initial state of the system is thus $|\psi (t=0)\rangle=|e\rangle\otimes|g\rangle\otimes|0\rangle$. We let the system evolve according to the master equation [Eq.\,\eqref{eq:master}] and calculate the populations of the respective qubits as a function of time and frequency $M$ of the acoustic drive as presented in Fig.\,\ref{fig:fig2} (a) for qubit 1 and in Fig.\,\ref{fig:fig2}(b) for qubit 2. As we tune $M$ the population dynamics qualitatively changes. In particular, when the integer multiple of $M$ matches the detuning $\delta_{12}$ between the bare qubit resonances, the dynamics feature Rabi oscillations, a signature of coherent cross talk between the qubits. This can be seen as dips (dark blue regions) in the color map in Fig.\,\ref{fig:fig2}(a) for $2M=\delta_{12}$ marking the minima of the population of qubit 1, which are correlated with the maxima (bright yellow regions) appearing in Fig.\,\ref{fig:fig2}(b) where the population of the second qubit is shown. For clarity we show cuts along the white-dashed lines shown in Fig.\,\ref{fig:fig2}(a,b) in Fig.\,\ref{fig:fig2}(c-e) for: (c) $2M=\delta_{12}$, (d) $M=\delta_{12}$, and (e) $M=\frac{3}{2}\delta_{12}$. Here populations of qubit 1 and qubit 2 are represented by the orange line and the blue-dashed line, respectively. Rabi oscillations are observed for both $2M=\delta_{12}$ and $M=\delta_{12}$, but these oscillations are markedly faster when $M=\delta_{12}$ [Fig.\,\ref{fig:fig2}(d)]. This is due to our choice of the driving amplitude $\mathcal{D}_1=\mathcal{D}_2\approx 0.92 M$ which leads to the optimized effective coupling as discussed in Section\,\ref{sec:model}. On the other hand in Fig.\,\ref{fig:fig2}(e) we show an example of a situation where the drive frequency (or its integer multiple) is mismatched with $\delta_{12}$. in this case the population of qubit 1 decays steadily to zero while qubit 2 remains in its ground state. 

\begin{figure*}
    \centering
    \includegraphics[scale=0.75]{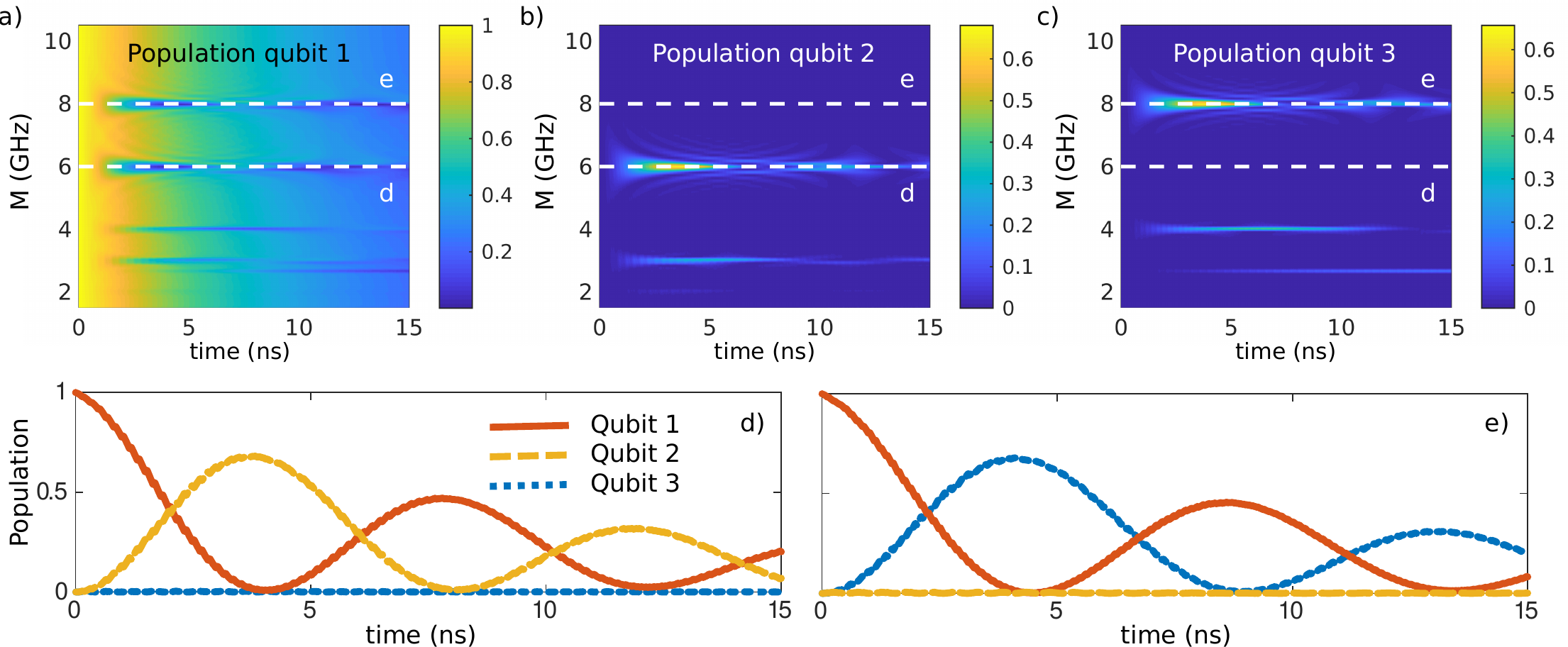}
    \caption{Selective qubit-qubit interaction demonstrated on a system of three qubits interacting with a single cavity mode. (a-c) Populations of (a) qubit 1, (b) qubit 2, and (c) qubit 3, as a function of time and the drive frequency $M$. The drive amplitude is adjusted such that $\mathcal{D}_i/M=\mathcal{D}/M\approx 0.92$ and the phases of the drive are chosen to be $\phi_2=\phi_3=\phi_1+\pi$ (i.e. qubit 1 interacts efficiently with qubit 2 and qubit 3, but qubits 2 and 3 do not mutually interact). Qubit 1 is initially in its excited state and the remaining parts of the system are in their ground state. If the integer multiple of the drive frequency matches with the difference of the excitation frequencies between qubit $i$ and qubit $j$, resonant interaction between these qubits is activated. This interaction selectively affects only the pair of qubits $i$ and $j$ if the difference $\delta_{ij}$ of the excitation frequencies is not equal to an integer multiple of $\delta_{ik}$ or $\delta_{jk}$. In (d,e) we show cuts along the lines drawn in (a,b) using orange line for population of qubit 1, yellow dashed line for qubit 2, and blue dotted line for qubit 3. (d) Rabi oscillations appear between qubit 1 and 2 for $M=\delta_{12}$ while qubit 3 is effectively decoupled from the dynamics. The reverse situation is shown in (e) for $M=\delta_{13}$.  }
    \label{fig:fig3}
\end{figure*}
Next we add into the system qubit 3 and perform an analogous calculation of population dynamics as for the qubit pair. The excited-state populations of the respective qubits 1-3 as a function of time and drive frequency $M$ are presented in Fig.\,\ref{fig:fig3} (a-c). As before, we assume that qubit 1 is initially excited and the rest of the system is in the ground state. As we tune the drive frequency we resonantly activate Rabi oscillations between qubit 1 and qubit 2 for $M=\delta_{12}$ and $2M=\delta_{12}$, but observe no population transfer to qubit 3 for this situation. Conversely, when $M=\delta_{13}$ and $2M=\delta_{13}$ is tuned, the coherent oscillation of populations occurs between qubit 1 and 3 while qubit 2 is not effected by this interaction. In Fig.\,\ref{fig:fig3}(d,e) we focus on the two first-order resonances achieved when $M=\delta_{12}$ [Fig.\,\ref{fig:fig3}(d)] and $M=\delta_{13}$ [Fig.\,\ref{fig:fig3}(e)] and plot the cuts along the white dashed lines marked in Fig.\,\ref{fig:fig3}(a-c). The detail confirms the selective character of the acoustically induced qubit-qubit interaction discussed above. 

We have thus shown that by tuning the frequency and amplitude of the external acoustic drive, we are indeed able to selectively dynamically switch on and off the interactions connecting a desired pair of qubits. This selectivity opens the possibility to steer the transfer of qubit states or of energy transfer through a network consisting of a larger number of optically and acoustically addressable quantum emitters.  


\section{Conclusions}
To summarize, we present a scheme allowing for selective acoustically induced photon-mediated coupling of qubits of different optical frequencies. If qubits are coupled dispersively to a common cavity mode, it is possible to apply an acoustic drive to dynamically dress the bare-qubit optical transitions and generate phonon sidebands in the qubit spectral response. These sidebands are detuned from the bare optical transition by an integer multiple of the drive frequency, $nM$, and can be used to trigger optical resonances between qubits if their respective frequency combs are overlapping. 

We have developed a theoretical model that can guide experimental design of devices that use the proposed scheme of selective acoustical control of photon-mediated qubit-qubit interactions. Devices based on diamond color centers as physical qubits could be a suitable platform for the experimental realization of our proposal. Such diamond-based devices would simultaneously allow for dispersive optical qubit-qubit coupling an acoustical control via the qubit hosting medium. Nevertheless, we point out that the scheme proposed here can be applied to any physical solid-state system if appropriate coupling and driving conditions can be engineered. For example, a similar method could be realized in hybrid diamond-piezomagnetic systems \cite{schuetz2015universaltransducers} where spin states of diamond defects could play the role of the physical qubit, and an acoustical resonator could be used instead of the optical cavity. Sideband generation via electrical driving has also been shown in silicon carbide quantum emitter systems \cite{Anderson2019-wo}. 

Looking forward, we expect that selective acoustical control of qubit-qubit interactions could be exploited in the design of fast and selective two-qubit gates and could enable spectral multiplexing as qubits of different colors can be entangled. One extension of the interaction shown in this work is a selective controlled-phase gate between ground-state qubits\cite{Asadi2019-lc}. This selective control would improve the speed of operations in systems based on solid-state quantum emitters and could be used to increase the density of physical qubits in solid-state quantum devices.  



\section*{Acknowledgements}
This material is based upon work supported by the U.S. Department of Energy, Office of Science, Basic Energy Sciences (BES), Materials Sciences and Engineering Division under FWP ERKCK47 and was performed at Harvard University.
The authors acknowledge the 'Photonics at Thermodynamic Limits' Energy Frontier Research Center funded by the U.S. Department of Energy, Office of Science, Office of Basic Energy Sciences under Award Number DE-SC0019140 that supported computational approaches used here. P. N. is a Moore Inventor Fellow supported by the Gordon and Betty Moore Foundation.

\appendix
\section{Emission spectrum of a single acoustically driven qubit}\label{app:emspec}
In Subsection\,\ref{subsubsec:adrive} of the main text we discuss how the emission properties of a single qubit are modified when the qubit is exposed to frequency modulation by an acoustic drive. This emission spectrum $s_{\rm e}(\omega)$ of a single acoustically driven qubit can be derived using the quantum regression theorem \cite{Breuer2003} starting from the definition
\begin{align}
    s_{\rm e}(\omega)\propto \lim_{T\to \infty}\frac{1}{T}\int_{-T/2}^{T/2}{\rm d}t\,\tilde{s}_{\rm e}(\omega, t),\label{seq:emspec}
\end{align}
with
\begin{align}
    \tilde{s}_{\rm e}(\omega, t)= \int_{-\infty}^\infty {\rm d}\tau \,\langle \sigma^\dagger (t)\sigma(t+\tau)\rangle e^{{\rm i}\omega\tau}.\label{seq:emspect}
\end{align}
In the definition of $s_{\rm e}(\omega)$ [Eq.\,\eqref{seq:emspec}] we have dropped the frequency-dependent prefactor $\propto \omega^4$ for simplicity. Equation\,\eqref{seq:emspect} can be evaluated using the result presented in Eq.\,\eqref{eq:intpicture}:
\begin{align}
    \sigma(t)=\sigma(0)e^{-{\rm i}{\omega_0}t}\sum_{n}J_n\left( \frac{\mathcal{D}}{M} \right)e^{-{\rm i}nMt}
\end{align}
and upon insertion into Eq.\,\eqref{seq:emspec} leads to:
\begin{align}
    s_{\rm e}&\propto\sum_{m,n} J_n\left( \frac{\mathcal{D}}{M} \right)J_m\left( \frac{\mathcal{D}}{M} \right)\delta(\omega-\omega_0-nM) \nonumber\\
    &\times \lim_{T\to\infty}\frac{1}{T}\int_{-T/2}^{T/2}{\rm d}t\,e^{-{\rm i}(n-m)Mt}\nonumber\\
    &=\sum_n \left [J_n\left( \frac{\mathcal{D}}{M} \right)\right]^2\delta(\omega-\omega_0-nM).\label{eq:specfin}
\end{align}
The last line of Eq.\,\eqref{eq:specfin} concludes that the spectral sidebands induced by the frequency modulation have amplitudes $\propto [J_n(\mathcal{D}/M)]^2$ (up to a frequency-dependent prefactor) as stated in the main text. 

\section{Breakdown of the secular approximation}\label{app:secular}
\begin{figure*}
    \centering
    \includegraphics[scale=0.75]{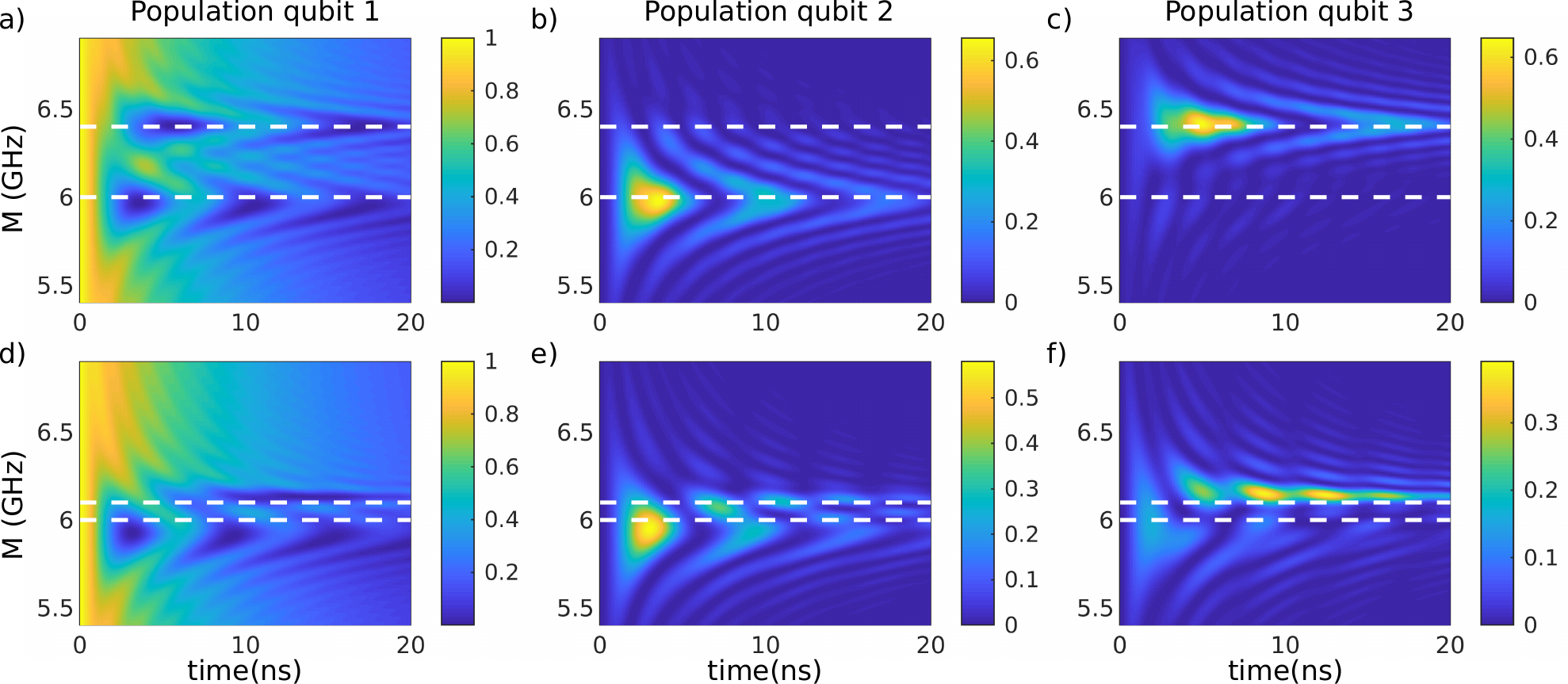}
    \caption{Breakdown of selectivity of the acoustically driven qubit-qubit interaction. Populations of qubit 1 (a,d), 2 (b,e), and 3 (c,f) as a function of time and frequency $M$ of the acoustic drive. We consider three coupled qubits of frequencies $\delta_1=2\pi\times 0$\,Hz, $\delta_2=2\pi\times 6$\,GHz, and (a-c) $\delta_3=2\pi\times 6.4$\,GHz, (d-f) $\delta_3=2\pi\times6.1$\,GHz. The qubits are driven with an equal amplitude $\mathcal{D}_i/M=\mathcal{D}/M=0.92$ but a different phase $\phi_2=\phi_3=\phi_1+\pi$. The frequency differences $\delta_{21}$ and $\delta_{31}$ are marked by the white dashed lines. We observe that the coherent exchange of population between qubit 1 and qubit 2 (qubit 3) is selective even for small differences between the target qubit frequencies as small as $\delta_{32}=2\pi\times 400$\,MHz (a-c), however, for $\delta_{32}=2\pi\times 100$\,MHz we see that despite the dominantly selective population transfer a considerable population is transferred to the off-resonant qubit (d-f). All the remaining parameters are given in Section\,\ref{sec:demonstration}.}
    \label{fig:figapp2}
\end{figure*}
The applicability of the method proposed relies strongly on the selectivity that allows for addressing individual qubit pairs. In this appendix we therefore examine the breakdown of the selective behavior as frequencies of two distinct target qubits (qubit 2 and 3) are brought closer to each other. this minimal separation of qubit energies that still preserves selectivity is closely associated with the breakdown of the secular approximation employed in the derivation of Eq.\,\eqref{eq:efham}. The minimal spectral separation of the qubit excitation frequencies should be $|\delta_{23}|\approx \mathcal{J}_{23}$. Using the parameters of Section\,\ref{sec:demonstration} we find $\mathcal{J}_{ij}\approx \mathcal{J}\approx 2\pi\times 100$\,MHz. In Fig.\,\ref{fig:figapp2} we show the populations of the three qubits as a function of time and drive frequency $M$ assuming the respective qubit frequencies $\delta_1=2\pi\times 0$\,Hz, $\delta_2=2\pi\times 6$\,GHz, and $\delta_3=2\pi\times 6.4$\,GHz in Fig.\,\ref{fig:figapp2}\,(a-c), or $\delta_3=2\pi\times6.1$\,GHz in Fig.\,\ref{fig:figapp2}\,(d-f). Qubit 1 is initially in the excited state and the rest of the system is in the ground state. The white dashed lines mark the spectral positions of the relevant detunings $\delta_{21}$ and $\delta_{31}$, respectively. In Fig.\,\ref{fig:figapp2}(a-c) we observe the mechanism of the acoustically driven selective resonant qubit-qubit coupling as described in Section\,\ref{sec:demonstration}, although some ripples appear in the population maps of qubit 2 and 3 even for off-resonant drive frequencies corresponding to the resonant interaction of qubit 1 with the other qubit (i.e. qubit 3 or 2, respectively). This situation is dramatically enhanced when the detuning is $\delta_{32}=2\pi\times 100$\,MHz ($\approx \mathcal{J}_{23}$), for which we observe the breakdown of the secular approximation and therefore of the selective behavior. We therefore conclude that the spectral separation of the qubit optical excitations should be $\approx 2\pi\times 100$\,MHz, although in practical situations where high-fidelity operations are required larger spectral separation may be needed.

\section{Physical parameters required for the optimal acoustic drive of diamond emitters}\label{app:physparam}
The ability to acoustically manipulate the qubit-qubit interaction relies on the susceptibility of the optically active qubit to strain that can be characterised via its deformation potential $D$. For diamond emitters this potential has been estimated as $D\sim 1$\,PHz \cite{golter2016optomechNV, lemonde2018phononnetworks}. Based on this value we discuss the parameters of an acoustic drive that need to be considered in practical experiments involving quantum emitters in diamond. We focus on the strain that deforms the crystal as a consequence of a propagating acoustic wave. As an example of a feasible acoustic drive we consider an acoustic plane wave of frequency $M=2\pi\times 5$\,GHz, characterised by a displacement $u=A_0 e^{{\rm i}k_{m}x}$, where $A_0$ is an amplitude,  $k=M/c_{\rm m}$, $c_{\rm m}=0.7\times 10^4$\,\si{\meter\cdot\second^{-1}} is the speed of sound in diamond \cite{lemonde2018phononnetworks}. We estimate the order of magnitude of the strain generated by such a wave as $|\varepsilon|\sim \left |\frac{\partial u}{\partial x}\right |=k_{\rm m}A_0$. For $M=2\pi\times 5$\,GHz, $k_{\rm m}\approx 3$\,\si{\micro\meter^{-1}} which corresponds to the wavelength of an acoustic wave of $\lambda_{\rm m}\approx 2$\,\si{\micro\meter}. The modulation amplitude $\mathcal{D}$ thus roughly becomes $\mathcal{D}=|\varepsilon|D\approx k_{\rm m} D A_0$ and we require that $\mathcal{D}/M\approx D A_0/c_{\rm m}\approx 1 $ to achieve the optimal coupling conditions. From the last relationship we estimate that the pumping amplitude in diamond should be $A_0\approx 10$\,pm (corresponding to the strain amplitude of $|\varepsilon|\approx 10^{-6}$). An optimal acoustic mode driving the optically active qubits would therefore be characterised by a wavelength of $\approx 1$\,\si{\micro\meter} and an oscillation amplitude of $\approx 10$\,\si{\pico\meter} (strain amplitude $|\varepsilon|\approx 10^{-6}$), which are values achievable in state-of-the-art experimental setups \cite{MacQuarrie2013mechanical, golter2016optomechNV,  chen2019engelphon, maity2019coherent}. Moreover, the wavelength corresponding to the acoustic drive is comparable to the wavelength of light at optical frequencies. Devices combining both the necessary optical and acoustical properties should therefore be within the experimental reach. 

\section{Effect of the qubit intrinsic damping and pure dephasing}\label{app:param}
\begin{figure*}
    \centering
    \includegraphics[scale=0.75]{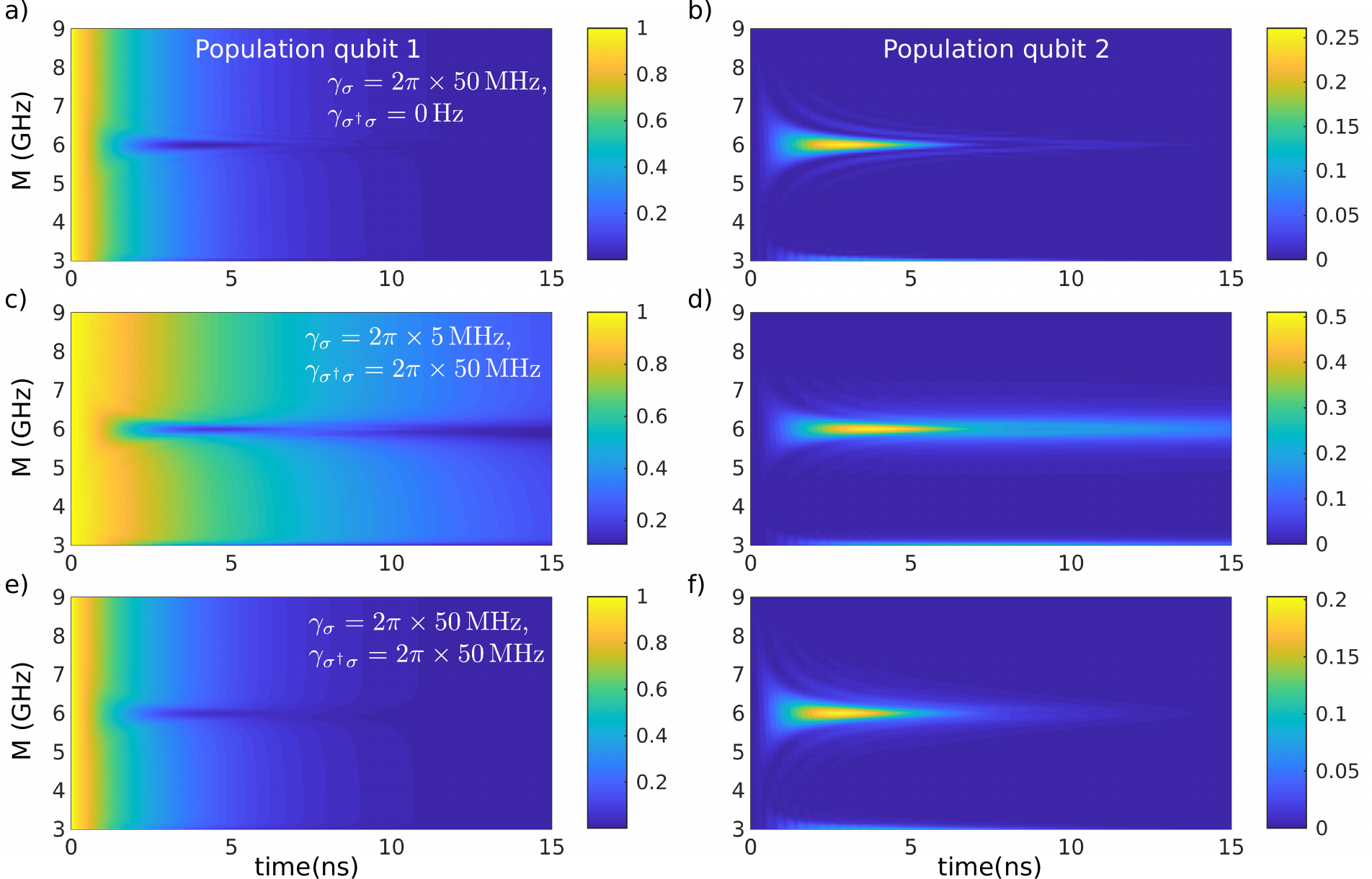}
    \caption{Influence of qubit intrinsic damping and pure dephasing on the acoustically-induced resonant interaction. The populations of qubit 1 and the corresponding populations of qubit 2 are plotted in (a,c,e) and (b,d,f), respectively, as a function of time and the frequency $M$ of the acoustic drive. The qubits are driven with an amplitude of $\mathcal{D}_i/M=\mathcal{D}/M=0.92$ and opposite relative phase $\phi_1=\phi_2+\pi$. The respective values of damping and dephasing are given in the inset and are identical for (a,b), (c,d), and (e,f), respectively. All the remaining parameters are given in Section\,\ref{sec:demonstration}.  }
    \label{fig:figapp1}
\end{figure*}
So far we have presented the system dynamics for a set of parameters that may corresponds to qubits represented by optically active diamond impurities coupled with a cavity in the form of a diamond photonic crystal. Nevertheless, we have assumed long intrinsic lifetime of the qubits and neglected effects of pure dephasing. In this appendix we show how pure dephasing and larger intrinsic damping influence the mechanism of selective acoustically driven resonant qubit-qubit coupling. 

To that end we consider the system of three qubits interacting with a single cavity mode and plot the resulting qubit populations as a function of time and drive frequency in Fig.\,\ref{fig:figapp1}. In Fig.\,\ref{fig:figapp1}(a,b) we increase the intrinsic damping to $\gamma_{\sigma}=2\pi\times 50$\,MHz, in Fig.\,\ref{fig:figapp1}(c,d) we consider small intrinsic losses $\gamma_{\sigma}=2\pi\times 5$\,MHz but large pure dephasing $\gamma_{\rm \sigma^\dagger\sigma}$, and finally in Fig.\,\ref{fig:figapp1}(e,f) both losses and dephasing are increased to $\gamma_{\sigma}=\gamma_{\sigma^\dagger\sigma}=2\pi\times 50$\,MHz. We see that when larger intrinsic losses are considered in the absence of pure dephasing, the amplitude of the first Rabi oscillation is quenched. In case that $\gamma_{\sigma^\dagger\sigma}=2\pi\times 50$\,MHz and the intrinsic losses are negligible, the exchange of population between qubit 1 and 2 persists until longer times, but the oscillatory character of the time evolution is suppressed. Similarly, the resonance of the qubit excitations is broadened. When both larger pure dephasing and larger intrinsic decay rate are considered, both the Rabi oscillations and the excited-state populations decay rapidly. This shortening of lifetime and coherence time is in all cases accompanied with a broadening of the spectral selectivity with respect to the drive frequency $M$.

\bibliography{biblio}

\newcommand{\noopsort}[1]{} \newcommand{\printfirst}[2]{#1}
  \newcommand{\singleletter}[1]{#1} \newcommand{\switchargs}[2]{#2#1}
\begin{thebibliography}{57}%
\makeatletter
\providecommand \@ifxundefined [1]{%
 \@ifx{#1\undefined}
}%
\providecommand \@ifnum [1]{%
 \ifnum #1\expandafter \@firstoftwo
 \else \expandafter \@secondoftwo
 \fi
}%
\providecommand \@ifx [1]{%
 \ifx #1\expandafter \@firstoftwo
 \else \expandafter \@secondoftwo
 \fi
}%
\providecommand \natexlab [1]{#1}%
\providecommand \enquote  [1]{``#1''}%
\providecommand \bibnamefont  [1]{#1}%
\providecommand \bibfnamefont [1]{#1}%
\providecommand \citenamefont [1]{#1}%
\providecommand \href@noop [0]{\@secondoftwo}%
\providecommand \href [0]{\begingroup \@sanitize@url \@href}%
\providecommand \@href[1]{\@@startlink{#1}\@@href}%
\providecommand \@@href[1]{\endgroup#1\@@endlink}%
\providecommand \@sanitize@url [0]{\catcode `\\12\catcode `\$12\catcode
  `\&12\catcode `\#12\catcode `\^12\catcode `\_12\catcode `\%12\relax}%
\providecommand \@@startlink[1]{}%
\providecommand \@@endlink[0]{}%
\providecommand \url  [0]{\begingroup\@sanitize@url \@url }%
\providecommand \@url [1]{\endgroup\@href {#1}{\urlprefix }}%
\providecommand \urlprefix  [0]{URL }%
\providecommand \Eprint [0]{\href }%
\providecommand \doibase [0]{http://dx.doi.org/}%
\providecommand \selectlanguage [0]{\@gobble}%
\providecommand \bibinfo  [0]{\@secondoftwo}%
\providecommand \bibfield  [0]{\@secondoftwo}%
\providecommand \translation [1]{[#1]}%
\providecommand \BibitemOpen [0]{}%
\providecommand \bibitemStop [0]{}%
\providecommand \bibitemNoStop [0]{.\EOS\space}%
\providecommand \EOS [0]{\spacefactor3000\relax}%
\providecommand \BibitemShut  [1]{\csname bibitem#1\endcsname}%
\let\auto@bib@innerbib\@empty
\bibitem [{\citenamefont {Awschalom}\ \emph {et~al.}(2018)\citenamefont
  {Awschalom}, \citenamefont {Hanson}, \citenamefont {Wrachtrup},\ and\
  \citenamefont {Zhou}}]{Awschalom2018-en}%
  \BibitemOpen
  \bibfield  {author} {\bibinfo {author} {\bibfnamefont {D.~D.}\ \bibnamefont
  {Awschalom}}, \bibinfo {author} {\bibfnamefont {R.}~\bibnamefont {Hanson}},
  \bibinfo {author} {\bibfnamefont {J.}~\bibnamefont {Wrachtrup}}, \ and\
  \bibinfo {author} {\bibfnamefont {B.~B.}\ \bibnamefont {Zhou}},\ }\href@noop
  {} {\bibfield  {journal} {\bibinfo  {journal} {Nat. Photonics}\ }\textbf
  {\bibinfo {volume} {12}},\ \bibinfo {pages} {516} (\bibinfo {year}
  {2018})}\BibitemShut {NoStop}%
\bibitem [{\citenamefont {Kimble}(2008)}]{Kimble2008-cu}%
  \BibitemOpen
  \bibfield  {author} {\bibinfo {author} {\bibfnamefont {H.~J.}\ \bibnamefont
  {Kimble}},\ }\href@noop {} {\bibfield  {journal} {\bibinfo  {journal}
  {Nature}\ }\textbf {\bibinfo {volume} {453}},\ \bibinfo {pages} {1023}
  (\bibinfo {year} {2008})}\BibitemShut {NoStop}%
\bibitem [{\citenamefont {Wehner}\ \emph {et~al.}(2018)\citenamefont {Wehner},
  \citenamefont {Elkouss},\ and\ \citenamefont {Hanson}}]{Wehner2018-wx}%
  \BibitemOpen
  \bibfield  {author} {\bibinfo {author} {\bibfnamefont {S.}~\bibnamefont
  {Wehner}}, \bibinfo {author} {\bibfnamefont {D.}~\bibnamefont {Elkouss}}, \
  and\ \bibinfo {author} {\bibfnamefont {R.}~\bibnamefont {Hanson}},\
  }\href@noop {} {\bibfield  {journal} {\bibinfo  {journal} {Science}\ }\textbf
  {\bibinfo {volume} {362}} (\bibinfo {year} {2018})}\BibitemShut {NoStop}%
\bibitem [{\citenamefont {Jiang}\ \emph {et~al.}(2007)\citenamefont {Jiang},
  \citenamefont {Taylor}, \citenamefont {Sorensen},\ and\ \citenamefont
  {Lukin}}]{Jiang2007-hr}%
  \BibitemOpen
  \bibfield  {author} {\bibinfo {author} {\bibfnamefont {L.}~\bibnamefont
  {Jiang}}, \bibinfo {author} {\bibfnamefont {J.~M.}\ \bibnamefont {Taylor}},
  \bibinfo {author} {\bibfnamefont {A.~S.}\ \bibnamefont {Sorensen}}, \ and\
  \bibinfo {author} {\bibfnamefont {M.~D.}\ \bibnamefont {Lukin}},\ }\href@noop
  {} {\bibfield  {journal} {\bibinfo  {journal} {Phys. Rev. A}\ }\textbf
  {\bibinfo {volume} {76}} (\bibinfo {year} {2007})}\BibitemShut {NoStop}%
\bibitem [{\citenamefont {Hucul}\ \emph {et~al.}(2015)\citenamefont {Hucul},
  \citenamefont {Inlek}, \citenamefont {Vittorini}, \citenamefont {Crocker},
  \citenamefont {Debnath}, \citenamefont {Clark},\ and\ \citenamefont
  {Monroe}}]{Hucul2015-td}%
  \BibitemOpen
  \bibfield  {author} {\bibinfo {author} {\bibfnamefont {D.}~\bibnamefont
  {Hucul}}, \bibinfo {author} {\bibfnamefont {I.~V.}\ \bibnamefont {Inlek}},
  \bibinfo {author} {\bibfnamefont {G.}~\bibnamefont {Vittorini}}, \bibinfo
  {author} {\bibfnamefont {C.}~\bibnamefont {Crocker}}, \bibinfo {author}
  {\bibfnamefont {S.}~\bibnamefont {Debnath}}, \bibinfo {author} {\bibfnamefont
  {S.~M.}\ \bibnamefont {Clark}}, \ and\ \bibinfo {author} {\bibfnamefont
  {C.}~\bibnamefont {Monroe}},\ }\href@noop {} {\bibfield  {journal} {\bibinfo
  {journal} {Nat. Phys.}\ }\textbf {\bibinfo {volume} {11}},\ \bibinfo {pages}
  {37} (\bibinfo {year} {2015})}\BibitemShut {NoStop}%
\bibitem [{\citenamefont {Dutt}\ \emph {et~al.}(2007)\citenamefont {Dutt},
  \citenamefont {Childress}, \citenamefont {Jiang}, \citenamefont {Togan},
  \citenamefont {Maze}, \citenamefont {Jelezko}, \citenamefont {Zibrov},
  \citenamefont {Hemmer},\ and\ \citenamefont {Lukin}}]{Dutt2007-tj}%
  \BibitemOpen
  \bibfield  {author} {\bibinfo {author} {\bibfnamefont {M.~V.~G.}\
  \bibnamefont {Dutt}}, \bibinfo {author} {\bibfnamefont {L.}~\bibnamefont
  {Childress}}, \bibinfo {author} {\bibfnamefont {L.}~\bibnamefont {Jiang}},
  \bibinfo {author} {\bibfnamefont {E.}~\bibnamefont {Togan}}, \bibinfo
  {author} {\bibfnamefont {J.}~\bibnamefont {Maze}}, \bibinfo {author}
  {\bibfnamefont {F.}~\bibnamefont {Jelezko}}, \bibinfo {author} {\bibfnamefont
  {A.~S.}\ \bibnamefont {Zibrov}}, \bibinfo {author} {\bibfnamefont {P.~R.}\
  \bibnamefont {Hemmer}}, \ and\ \bibinfo {author} {\bibfnamefont {M.~D.}\
  \bibnamefont {Lukin}},\ }\href@noop {} {\bibfield  {journal} {\bibinfo
  {journal} {Science}\ }\textbf {\bibinfo {volume} {316}},\ \bibinfo {pages}
  {1312} (\bibinfo {year} {2007})}\BibitemShut {NoStop}%
\bibitem [{\citenamefont {Robledo}\ \emph {et~al.}(2011)\citenamefont
  {Robledo}, \citenamefont {Childress}, \citenamefont {Bernien}, \citenamefont
  {Hensen}, \citenamefont {Alkemade},\ and\ \citenamefont
  {Hanson}}]{Robledo2011-fw}%
  \BibitemOpen
  \bibfield  {author} {\bibinfo {author} {\bibfnamefont {L.}~\bibnamefont
  {Robledo}}, \bibinfo {author} {\bibfnamefont {L.}~\bibnamefont {Childress}},
  \bibinfo {author} {\bibfnamefont {H.}~\bibnamefont {Bernien}}, \bibinfo
  {author} {\bibfnamefont {B.}~\bibnamefont {Hensen}}, \bibinfo {author}
  {\bibfnamefont {P.~F.~A.}\ \bibnamefont {Alkemade}}, \ and\ \bibinfo {author}
  {\bibfnamefont {R.}~\bibnamefont {Hanson}},\ }\href@noop {} {\bibfield
  {journal} {\bibinfo  {journal} {Nature}\ }\textbf {\bibinfo {volume} {477}},\
  \bibinfo {pages} {574} (\bibinfo {year} {2011})}\BibitemShut {NoStop}%
\bibitem [{\citenamefont {Dolde}\ \emph {et~al.}(2013)\citenamefont {Dolde},
  \citenamefont {Jakobi}, \citenamefont {Naydenov}, \citenamefont {Zhao},
  \citenamefont {Pezzagna}, \citenamefont {Trautmann}, \citenamefont {Meijer},
  \citenamefont {Neumann}, \citenamefont {Jelezko},\ and\ \citenamefont
  {Wrachtrup}}]{Dolde2013-vh}%
  \BibitemOpen
  \bibfield  {author} {\bibinfo {author} {\bibfnamefont {F.}~\bibnamefont
  {Dolde}}, \bibinfo {author} {\bibfnamefont {I.}~\bibnamefont {Jakobi}},
  \bibinfo {author} {\bibfnamefont {B.}~\bibnamefont {Naydenov}}, \bibinfo
  {author} {\bibfnamefont {N.}~\bibnamefont {Zhao}}, \bibinfo {author}
  {\bibfnamefont {S.}~\bibnamefont {Pezzagna}}, \bibinfo {author}
  {\bibfnamefont {C.}~\bibnamefont {Trautmann}}, \bibinfo {author}
  {\bibfnamefont {J.}~\bibnamefont {Meijer}}, \bibinfo {author} {\bibfnamefont
  {P.}~\bibnamefont {Neumann}}, \bibinfo {author} {\bibfnamefont
  {F.}~\bibnamefont {Jelezko}}, \ and\ \bibinfo {author} {\bibfnamefont
  {J.}~\bibnamefont {Wrachtrup}},\ }\href@noop {} {\bibfield  {journal}
  {\bibinfo  {journal} {Nat. Phys.}\ }\textbf {\bibinfo {volume} {8}},\
  \bibinfo {pages} {1} (\bibinfo {year} {2013})}\BibitemShut {NoStop}%
\bibitem [{\citenamefont {Turchette}\ \emph {et~al.}(1998)\citenamefont
  {Turchette}, \citenamefont {Wood}, \citenamefont {King}, \citenamefont
  {Myatt}, \citenamefont {Leibfried}, \citenamefont {Itano}, \citenamefont
  {Monroe},\ and\ \citenamefont {Wineland}}]{Turchette1998-kf}%
  \BibitemOpen
  \bibfield  {author} {\bibinfo {author} {\bibfnamefont {Q.~A.}\ \bibnamefont
  {Turchette}}, \bibinfo {author} {\bibfnamefont {C.~S.}\ \bibnamefont {Wood}},
  \bibinfo {author} {\bibfnamefont {B.~E.}\ \bibnamefont {King}}, \bibinfo
  {author} {\bibfnamefont {C.~J.}\ \bibnamefont {Myatt}}, \bibinfo {author}
  {\bibfnamefont {D.}~\bibnamefont {Leibfried}}, \bibinfo {author}
  {\bibfnamefont {W.~M.}\ \bibnamefont {Itano}}, \bibinfo {author}
  {\bibfnamefont {C.}~\bibnamefont {Monroe}}, \ and\ \bibinfo {author}
  {\bibfnamefont {D.~J.}\ \bibnamefont {Wineland}},\ }\href@noop {} {\bibfield
  {journal} {\bibinfo  {journal} {Phys. Rev. Lett.}\ }\textbf {\bibinfo
  {volume} {81}},\ \bibinfo {pages} {3631} (\bibinfo {year}
  {1998})}\BibitemShut {NoStop}%
\bibitem [{\citenamefont {Monroe}\ and\ \citenamefont
  {Kim}(2013)}]{Monroe2013-mq}%
  \BibitemOpen
  \bibfield  {author} {\bibinfo {author} {\bibfnamefont {C.}~\bibnamefont
  {Monroe}}\ and\ \bibinfo {author} {\bibfnamefont {J.}~\bibnamefont {Kim}},\
  }\href@noop {} {\bibfield  {journal} {\bibinfo  {journal} {Science}\ }\textbf
  {\bibinfo {volume} {339}},\ \bibinfo {pages} {1164} (\bibinfo {year}
  {2013})}\BibitemShut {NoStop}%
\bibitem [{\citenamefont {Welte}\ \emph {et~al.}(2018)\citenamefont {Welte},
  \citenamefont {Hacker}, \citenamefont {Daiss}, \citenamefont {Ritter},\ and\
  \citenamefont {Rempe}}]{Welte2018-us}%
  \BibitemOpen
  \bibfield  {author} {\bibinfo {author} {\bibfnamefont {S.}~\bibnamefont
  {Welte}}, \bibinfo {author} {\bibfnamefont {B.}~\bibnamefont {Hacker}},
  \bibinfo {author} {\bibfnamefont {S.}~\bibnamefont {Daiss}}, \bibinfo
  {author} {\bibfnamefont {S.}~\bibnamefont {Ritter}}, \ and\ \bibinfo {author}
  {\bibfnamefont {G.}~\bibnamefont {Rempe}},\ }\href@noop {} {\bibfield
  {journal} {\bibinfo  {journal} {Phys. Rev. X}\ }\textbf {\bibinfo {volume}
  {8}},\ \bibinfo {pages} {011018} (\bibinfo {year} {2018})}\BibitemShut
  {NoStop}%
\bibitem [{\citenamefont {Evans}\ \emph {et~al.}(2018)\citenamefont {Evans},
  \citenamefont {Bhaskar}, \citenamefont {Sukachev}, \citenamefont {Nguyen},
  \citenamefont {Sipahigil}, \citenamefont {Burek}, \citenamefont {Machielse},
  \citenamefont {Zhang}, \citenamefont {Zibrov}, \citenamefont {Bielejec},
  \citenamefont {Park}, \citenamefont {Lon{\v c}ar},\ and\ \citenamefont
  {Lukin}}]{Evans2018-vh}%
  \BibitemOpen
  \bibfield  {author} {\bibinfo {author} {\bibfnamefont {R.~E.}\ \bibnamefont
  {Evans}}, \bibinfo {author} {\bibfnamefont {M.~K.}\ \bibnamefont {Bhaskar}},
  \bibinfo {author} {\bibfnamefont {D.~D.}\ \bibnamefont {Sukachev}}, \bibinfo
  {author} {\bibfnamefont {C.~T.}\ \bibnamefont {Nguyen}}, \bibinfo {author}
  {\bibfnamefont {A.}~\bibnamefont {Sipahigil}}, \bibinfo {author}
  {\bibfnamefont {M.~J.}\ \bibnamefont {Burek}}, \bibinfo {author}
  {\bibfnamefont {B.}~\bibnamefont {Machielse}}, \bibinfo {author}
  {\bibfnamefont {G.~H.}\ \bibnamefont {Zhang}}, \bibinfo {author}
  {\bibfnamefont {A.~S.}\ \bibnamefont {Zibrov}}, \bibinfo {author}
  {\bibfnamefont {E.}~\bibnamefont {Bielejec}}, \bibinfo {author}
  {\bibfnamefont {H.}~\bibnamefont {Park}}, \bibinfo {author} {\bibfnamefont
  {M.}~\bibnamefont {Lon{\v c}ar}}, \ and\ \bibinfo {author} {\bibfnamefont
  {M.~D.}\ \bibnamefont {Lukin}},\ }\href@noop {} {\bibfield  {journal}
  {\bibinfo  {journal} {Science}\ }\textbf {\bibinfo {volume} {362}},\ \bibinfo
  {pages} {662} (\bibinfo {year} {2018})}\BibitemShut {NoStop}%
\bibitem [{\citenamefont {Gavartin}\ \emph {et~al.}(2011)\citenamefont
  {Gavartin}, \citenamefont {Braive}, \citenamefont {Sagnes}, \citenamefont
  {Arcizet}, \citenamefont {Beveratos}, \citenamefont {Kippenberg},\ and\
  \citenamefont {Robert-Philip}}]{gavartin2011optomechcouple}%
  \BibitemOpen
  \bibfield  {author} {\bibinfo {author} {\bibfnamefont {E.}~\bibnamefont
  {Gavartin}}, \bibinfo {author} {\bibfnamefont {R.}~\bibnamefont {Braive}},
  \bibinfo {author} {\bibfnamefont {I.}~\bibnamefont {Sagnes}}, \bibinfo
  {author} {\bibfnamefont {O.}~\bibnamefont {Arcizet}}, \bibinfo {author}
  {\bibfnamefont {A.}~\bibnamefont {Beveratos}}, \bibinfo {author}
  {\bibfnamefont {T.~J.}\ \bibnamefont {Kippenberg}}, \ and\ \bibinfo {author}
  {\bibfnamefont {I.}~\bibnamefont {Robert-Philip}},\ }\href {\doibase
  10.1103/PhysRevLett.106.203902} {\bibfield  {journal} {\bibinfo  {journal}
  {Phys. Rev. Lett.}\ }\textbf {\bibinfo {volume} {106}},\ \bibinfo {pages}
  {203902} (\bibinfo {year} {2011})}\BibitemShut {NoStop}%
\bibitem [{\citenamefont {Aspelmeyer}\ \emph {et~al.}(2012)\citenamefont
  {Aspelmeyer}, \citenamefont {Meystre},\ and\ \citenamefont
  {Schwab}}]{Aspelmeyer2012-sy}%
  \BibitemOpen
  \bibfield  {author} {\bibinfo {author} {\bibfnamefont {M.}~\bibnamefont
  {Aspelmeyer}}, \bibinfo {author} {\bibfnamefont {P.}~\bibnamefont {Meystre}},
  \ and\ \bibinfo {author} {\bibfnamefont {K.}~\bibnamefont {Schwab}},\
  }\href@noop {} {\bibfield  {journal} {\bibinfo  {journal} {Phys. Today}\
  }\textbf {\bibinfo {volume} {65}},\ \bibinfo {pages} {29} (\bibinfo {year}
  {2012})}\BibitemShut {NoStop}%
\bibitem [{\citenamefont {Galland}\ \emph {et~al.}(2014)\citenamefont
  {Galland}, \citenamefont {Sangouard}, \citenamefont {Piro}, \citenamefont
  {Gisin},\ and\ \citenamefont {Kippenberg}}]{Galland2014-cl}%
  \BibitemOpen
  \bibfield  {author} {\bibinfo {author} {\bibfnamefont {C.}~\bibnamefont
  {Galland}}, \bibinfo {author} {\bibfnamefont {N.}~\bibnamefont {Sangouard}},
  \bibinfo {author} {\bibfnamefont {N.}~\bibnamefont {Piro}}, \bibinfo {author}
  {\bibfnamefont {N.}~\bibnamefont {Gisin}}, \ and\ \bibinfo {author}
  {\bibfnamefont {T.~J.}\ \bibnamefont {Kippenberg}},\ }\href@noop {}
  {\bibfield  {journal} {\bibinfo  {journal} {Phys. Rev. Lett.}\ }\textbf
  {\bibinfo {volume} {112}},\ \bibinfo {pages} {143602} (\bibinfo {year}
  {2014})}\BibitemShut {NoStop}%
\bibitem [{\citenamefont {Safavi-Naeini}\ \emph {et~al.}(2014)\citenamefont
  {Safavi-Naeini}, \citenamefont {Hill}, \citenamefont {Meenehan},
  \citenamefont {Chan}, \citenamefont {Gr{\"o}blacher},\ and\ \citenamefont
  {Painter}}]{Safavi-Naeini2014-af}%
  \BibitemOpen
  \bibfield  {author} {\bibinfo {author} {\bibfnamefont {A.~H.}\ \bibnamefont
  {Safavi-Naeini}}, \bibinfo {author} {\bibfnamefont {J.~T.}\ \bibnamefont
  {Hill}}, \bibinfo {author} {\bibfnamefont {S.}~\bibnamefont {Meenehan}},
  \bibinfo {author} {\bibfnamefont {J.}~\bibnamefont {Chan}}, \bibinfo {author}
  {\bibfnamefont {S.}~\bibnamefont {Gr{\"o}blacher}}, \ and\ \bibinfo {author}
  {\bibfnamefont {O.}~\bibnamefont {Painter}},\ }\href@noop {} {\bibfield
  {journal} {\bibinfo  {journal} {Phys. Rev. Lett.}\ }\textbf {\bibinfo
  {volume} {112}},\ \bibinfo {pages} {153603} (\bibinfo {year}
  {2014})}\BibitemShut {NoStop}%
\bibitem [{\citenamefont {Fang}\ \emph {et~al.}(2016)\citenamefont {Fang},
  \citenamefont {Matheny}, \citenamefont {Luan},\ and\ \citenamefont
  {Painter}}]{fang2016optical}%
  \BibitemOpen
  \bibfield  {author} {\bibinfo {author} {\bibfnamefont {K.}~\bibnamefont
  {Fang}}, \bibinfo {author} {\bibfnamefont {M.~H.}\ \bibnamefont {Matheny}},
  \bibinfo {author} {\bibfnamefont {X.}~\bibnamefont {Luan}}, \ and\ \bibinfo
  {author} {\bibfnamefont {O.}~\bibnamefont {Painter}},\ }\href@noop {}
  {\bibfield  {journal} {\bibinfo  {journal} {Nat. Photon.}\ }\textbf {\bibinfo
  {volume} {10}},\ \bibinfo {pages} {489–496} (\bibinfo {year}
  {2016})}\BibitemShut {NoStop}%
\bibitem [{\citenamefont {Golter}\ \emph {et~al.}(2016)\citenamefont {Golter},
  \citenamefont {Oo}, \citenamefont {Amezcua}, \citenamefont {Stewart},\ and\
  \citenamefont {Wang}}]{golter2016optomechNV}%
  \BibitemOpen
  \bibfield  {author} {\bibinfo {author} {\bibfnamefont {D.~A.}\ \bibnamefont
  {Golter}}, \bibinfo {author} {\bibfnamefont {T.}~\bibnamefont {Oo}}, \bibinfo
  {author} {\bibfnamefont {M.}~\bibnamefont {Amezcua}}, \bibinfo {author}
  {\bibfnamefont {K.~A.}\ \bibnamefont {Stewart}}, \ and\ \bibinfo {author}
  {\bibfnamefont {H.}~\bibnamefont {Wang}},\ }\href {\doibase
  10.1103/PhysRevLett.116.143602} {\bibfield  {journal} {\bibinfo  {journal}
  {Phys. Rev. Lett.}\ }\textbf {\bibinfo {volume} {116}},\ \bibinfo {pages}
  {143602} (\bibinfo {year} {2016})}\BibitemShut {NoStop}%
\bibitem [{\citenamefont {Burek}\ \emph {et~al.}(2016)\citenamefont {Burek},
  \citenamefont {Cohen}, \citenamefont {Meenehan}, \citenamefont {El-Sawah},
  \citenamefont {Chia}, \citenamefont {Ruelle}, \citenamefont {Meesala},
  \citenamefont {Rochman}, \citenamefont {Atikian}, \citenamefont {Markham},
  \citenamefont {Twitchen}, \citenamefont {Lukin}, \citenamefont {Painter},\
  and\ \citenamefont {Lon\v{c}ar}}]{burek2016diamondoptomech}%
  \BibitemOpen
  \bibfield  {author} {\bibinfo {author} {\bibfnamefont {M.~J.}\ \bibnamefont
  {Burek}}, \bibinfo {author} {\bibfnamefont {J.~D.}\ \bibnamefont {Cohen}},
  \bibinfo {author} {\bibfnamefont {S.~M.}\ \bibnamefont {Meenehan}}, \bibinfo
  {author} {\bibfnamefont {N.}~\bibnamefont {El-Sawah}}, \bibinfo {author}
  {\bibfnamefont {C.}~\bibnamefont {Chia}}, \bibinfo {author} {\bibfnamefont
  {T.}~\bibnamefont {Ruelle}}, \bibinfo {author} {\bibfnamefont
  {S.}~\bibnamefont {Meesala}}, \bibinfo {author} {\bibfnamefont
  {J.}~\bibnamefont {Rochman}}, \bibinfo {author} {\bibfnamefont {H.~A.}\
  \bibnamefont {Atikian}}, \bibinfo {author} {\bibfnamefont {M.}~\bibnamefont
  {Markham}}, \bibinfo {author} {\bibfnamefont {D.~J.}\ \bibnamefont
  {Twitchen}}, \bibinfo {author} {\bibfnamefont {M.~D.}\ \bibnamefont {Lukin}},
  \bibinfo {author} {\bibfnamefont {O.}~\bibnamefont {Painter}}, \ and\
  \bibinfo {author} {\bibfnamefont {M.}~\bibnamefont {Lon\v{c}ar}},\ }\href
  {\doibase 10.1364/OPTICA.3.001404} {\bibfield  {journal} {\bibinfo  {journal}
  {Optica}\ }\textbf {\bibinfo {volume} {3}},\ \bibinfo {pages} {1404}
  (\bibinfo {year} {2016})}\BibitemShut {NoStop}%
\bibitem [{\citenamefont {Lemonde}\ \emph {et~al.}(2018)\citenamefont
  {Lemonde}, \citenamefont {Meesala}, \citenamefont {Sipahigil}, \citenamefont
  {Schuetz}, \citenamefont {Lukin}, \citenamefont {Loncar},\ and\ \citenamefont
  {Rabl}}]{lemonde2018phononnetworks}%
  \BibitemOpen
  \bibfield  {author} {\bibinfo {author} {\bibfnamefont {M.-A.}\ \bibnamefont
  {Lemonde}}, \bibinfo {author} {\bibfnamefont {S.}~\bibnamefont {Meesala}},
  \bibinfo {author} {\bibfnamefont {A.}~\bibnamefont {Sipahigil}}, \bibinfo
  {author} {\bibfnamefont {M.~J.~A.}\ \bibnamefont {Schuetz}}, \bibinfo
  {author} {\bibfnamefont {M.~D.}\ \bibnamefont {Lukin}}, \bibinfo {author}
  {\bibfnamefont {M.}~\bibnamefont {Loncar}}, \ and\ \bibinfo {author}
  {\bibfnamefont {P.}~\bibnamefont {Rabl}},\ }\href {\doibase
  10.1103/PhysRevLett.120.213603} {\bibfield  {journal} {\bibinfo  {journal}
  {Phys. Rev. Lett.}\ }\textbf {\bibinfo {volume} {120}},\ \bibinfo {pages}
  {213603} (\bibinfo {year} {2018})}\BibitemShut {NoStop}%
\bibitem [{\citenamefont {Moores}\ \emph {et~al.}(2018)\citenamefont {Moores},
  \citenamefont {Sletten}, \citenamefont {Viennot},\ and\ \citenamefont
  {Lehnert}}]{moores2018quantumacoustic}%
  \BibitemOpen
  \bibfield  {author} {\bibinfo {author} {\bibfnamefont {B.~A.}\ \bibnamefont
  {Moores}}, \bibinfo {author} {\bibfnamefont {L.~R.}\ \bibnamefont {Sletten}},
  \bibinfo {author} {\bibfnamefont {J.~J.}\ \bibnamefont {Viennot}}, \ and\
  \bibinfo {author} {\bibfnamefont {K.~W.}\ \bibnamefont {Lehnert}},\ }\href
  {\doibase 10.1103/PhysRevLett.120.227701} {\bibfield  {journal} {\bibinfo
  {journal} {Phys. Rev. Lett.}\ }\textbf {\bibinfo {volume} {120}},\ \bibinfo
  {pages} {227701} (\bibinfo {year} {2018})}\BibitemShut {NoStop}%
\bibitem [{\citenamefont {Whiteley}\ \emph {et~al.}(2019)\citenamefont
  {Whiteley}, \citenamefont {Wolfowicz}, \citenamefont {Anderson},
  \citenamefont {Bourassa}, \citenamefont {Ma}, \citenamefont {Ye},
  \citenamefont {Koolstra}, \citenamefont {Satzinger}, \citenamefont {Holt},
  \citenamefont {Heremans} \emph {et~al.}}]{whiteley2019spin}%
  \BibitemOpen
  \bibfield  {author} {\bibinfo {author} {\bibfnamefont {S.~J.}\ \bibnamefont
  {Whiteley}}, \bibinfo {author} {\bibfnamefont {G.}~\bibnamefont {Wolfowicz}},
  \bibinfo {author} {\bibfnamefont {C.~P.}\ \bibnamefont {Anderson}}, \bibinfo
  {author} {\bibfnamefont {A.}~\bibnamefont {Bourassa}}, \bibinfo {author}
  {\bibfnamefont {H.}~\bibnamefont {Ma}}, \bibinfo {author} {\bibfnamefont
  {M.}~\bibnamefont {Ye}}, \bibinfo {author} {\bibfnamefont {G.}~\bibnamefont
  {Koolstra}}, \bibinfo {author} {\bibfnamefont {K.~J.}\ \bibnamefont
  {Satzinger}}, \bibinfo {author} {\bibfnamefont {M.~V.}\ \bibnamefont {Holt}},
  \bibinfo {author} {\bibfnamefont {F.~J.}\ \bibnamefont {Heremans}},  \emph
  {et~al.},\ }\href@noop {} {\bibfield  {journal} {\bibinfo  {journal} {Nat.
  Phys.}\ }\textbf {\bibinfo {volume} {15}},\ \bibinfo {pages} {490} (\bibinfo
  {year} {2019})}\BibitemShut {NoStop}%
\bibitem [{\citenamefont {MacCabe}\ \emph {et~al.}(2019)\citenamefont
  {MacCabe}, \citenamefont {Ren}, \citenamefont {Luo}, \citenamefont {Cohen},
  \citenamefont {Zhou}, \citenamefont {Sipahigil}, \citenamefont
  {Mirhosseini},\ and\ \citenamefont {Painter}}]{maccabe2019phononic}%
  \BibitemOpen
  \bibfield  {author} {\bibinfo {author} {\bibfnamefont {G.~S.}\ \bibnamefont
  {MacCabe}}, \bibinfo {author} {\bibfnamefont {H.}~\bibnamefont {Ren}},
  \bibinfo {author} {\bibfnamefont {J.}~\bibnamefont {Luo}}, \bibinfo {author}
  {\bibfnamefont {J.~D.}\ \bibnamefont {Cohen}}, \bibinfo {author}
  {\bibfnamefont {H.}~\bibnamefont {Zhou}}, \bibinfo {author} {\bibfnamefont
  {A.}~\bibnamefont {Sipahigil}}, \bibinfo {author} {\bibfnamefont
  {M.}~\bibnamefont {Mirhosseini}}, \ and\ \bibinfo {author} {\bibfnamefont
  {O.}~\bibnamefont {Painter}},\ }\href@noop {} {\bibfield  {journal} {\bibinfo
   {journal} {arXiv preprint arXiv:1901.04129}\ } (\bibinfo {year}
  {2019})}\BibitemShut {NoStop}%
\bibitem [{\citenamefont {Ovartchaiyapong}\ \emph {et~al.}(2014)\citenamefont
  {Ovartchaiyapong}, \citenamefont {Lee}, \citenamefont {Myers},\ and\
  \citenamefont {Jayich}}]{ovartchaiyapong2014dynamic}%
  \BibitemOpen
  \bibfield  {author} {\bibinfo {author} {\bibfnamefont {P.}~\bibnamefont
  {Ovartchaiyapong}}, \bibinfo {author} {\bibfnamefont {K.~W.}\ \bibnamefont
  {Lee}}, \bibinfo {author} {\bibfnamefont {B.~A.}\ \bibnamefont {Myers}}, \
  and\ \bibinfo {author} {\bibfnamefont {A.~C.~B.}\ \bibnamefont {Jayich}},\
  }\href@noop {} {\bibfield  {journal} {\bibinfo  {journal} {Nat. Commun.}\
  }\textbf {\bibinfo {volume} {5}},\ \bibinfo {pages} {4429} (\bibinfo {year}
  {2014})}\BibitemShut {NoStop}%
\bibitem [{\citenamefont {Schuetz}\ \emph {et~al.}(2015)\citenamefont
  {Schuetz}, \citenamefont {Kessler}, \citenamefont {Giedke}, \citenamefont
  {Vandersypen}, \citenamefont {Lukin},\ and\ \citenamefont
  {Cirac}}]{schuetz2015universaltransducers}%
  \BibitemOpen
  \bibfield  {author} {\bibinfo {author} {\bibfnamefont {M.~J.~A.}\
  \bibnamefont {Schuetz}}, \bibinfo {author} {\bibfnamefont {E.~M.}\
  \bibnamefont {Kessler}}, \bibinfo {author} {\bibfnamefont {G.}~\bibnamefont
  {Giedke}}, \bibinfo {author} {\bibfnamefont {L.~M.~K.}\ \bibnamefont
  {Vandersypen}}, \bibinfo {author} {\bibfnamefont {M.~D.}\ \bibnamefont
  {Lukin}}, \ and\ \bibinfo {author} {\bibfnamefont {J.~I.}\ \bibnamefont
  {Cirac}},\ }\href {\doibase 10.1103/PhysRevX.5.031031} {\bibfield  {journal}
  {\bibinfo  {journal} {Phys. Rev. X}\ }\textbf {\bibinfo {volume} {5}},\
  \bibinfo {pages} {031031} (\bibinfo {year} {2015})}\BibitemShut {NoStop}%
\bibitem [{\citenamefont {MacQuarrie}\ \emph {et~al.}(2015)\citenamefont
  {MacQuarrie}, \citenamefont {Gosavi}, \citenamefont {Bhave},\ and\
  \citenamefont {Fuchs}}]{macquarrie2015contdecoupling}%
  \BibitemOpen
  \bibfield  {author} {\bibinfo {author} {\bibfnamefont {E.~R.}\ \bibnamefont
  {MacQuarrie}}, \bibinfo {author} {\bibfnamefont {T.~A.}\ \bibnamefont
  {Gosavi}}, \bibinfo {author} {\bibfnamefont {S.~A.}\ \bibnamefont {Bhave}}, \
  and\ \bibinfo {author} {\bibfnamefont {G.~D.}\ \bibnamefont {Fuchs}},\ }\href
  {\doibase 10.1103/PhysRevB.92.224419} {\bibfield  {journal} {\bibinfo
  {journal} {Phys. Rev. B}\ }\textbf {\bibinfo {volume} {92}},\ \bibinfo
  {pages} {224419} (\bibinfo {year} {2015})}\BibitemShut {NoStop}%
\bibitem [{\citenamefont {Jahnke}\ \emph {et~al.}(2015)\citenamefont {Jahnke},
  \citenamefont {Sipahigil}, \citenamefont {Binder}, \citenamefont {Doherty},
  \citenamefont {Metsch}, \citenamefont {Rogers}, \citenamefont {Manson},
  \citenamefont {Lukin},\ and\ \citenamefont {Jelezko}}]{Jahnke2015elphonon}%
  \BibitemOpen
  \bibfield  {author} {\bibinfo {author} {\bibfnamefont {K.~D.}\ \bibnamefont
  {Jahnke}}, \bibinfo {author} {\bibfnamefont {A.}~\bibnamefont {Sipahigil}},
  \bibinfo {author} {\bibfnamefont {J.~M.}\ \bibnamefont {Binder}}, \bibinfo
  {author} {\bibfnamefont {M.~W.}\ \bibnamefont {Doherty}}, \bibinfo {author}
  {\bibfnamefont {M.}~\bibnamefont {Metsch}}, \bibinfo {author} {\bibfnamefont
  {L.~J.}\ \bibnamefont {Rogers}}, \bibinfo {author} {\bibfnamefont {N.~B.}\
  \bibnamefont {Manson}}, \bibinfo {author} {\bibfnamefont {M.~D.}\
  \bibnamefont {Lukin}}, \ and\ \bibinfo {author} {\bibfnamefont
  {F.}~\bibnamefont {Jelezko}},\ }\href {\doibase
  10.1088/1367-2630/17/4/043011} {\bibfield  {journal} {\bibinfo  {journal}
  {New J. Phys.}\ }\textbf {\bibinfo {volume} {17}},\ \bibinfo {pages} {043011}
  (\bibinfo {year} {2015})}\BibitemShut {NoStop}%
\bibitem [{\citenamefont {Chen}\ \emph {et~al.}(2019)\citenamefont {Chen},
  \citenamefont {Opondo}, \citenamefont {Jiang}, \citenamefont {MacQuarrie},
  \citenamefont {Daveau}, \citenamefont {Bhave},\ and\ \citenamefont
  {Fuchs}}]{chen2019engelphon}%
  \BibitemOpen
  \bibfield  {author} {\bibinfo {author} {\bibfnamefont {H.}~\bibnamefont
  {Chen}}, \bibinfo {author} {\bibfnamefont {N.~F.}\ \bibnamefont {Opondo}},
  \bibinfo {author} {\bibfnamefont {B.}~\bibnamefont {Jiang}}, \bibinfo
  {author} {\bibfnamefont {E.~R.}\ \bibnamefont {MacQuarrie}}, \bibinfo
  {author} {\bibfnamefont {R.~S.}\ \bibnamefont {Daveau}}, \bibinfo {author}
  {\bibfnamefont {S.~A.}\ \bibnamefont {Bhave}}, \ and\ \bibinfo {author}
  {\bibfnamefont {G.~D.}\ \bibnamefont {Fuchs}},\ }\href {\doibase
  10.1021/acs.nanolett.9b02430} {\bibfield  {journal} {\bibinfo  {journal}
  {Nano Letters}\ }\textbf {\bibinfo {volume} {19}},\ \bibinfo {pages} {7021}
  (\bibinfo {year} {2019})},\ \bibinfo {note} {pMID: 31498998}\BibitemShut
  {NoStop}%
\bibitem [{\citenamefont {Li}\ \emph {et~al.}(2019)\citenamefont {Li},
  \citenamefont {Kuzyk},\ and\ \citenamefont {Wang}}]{li2019honeycomb}%
  \BibitemOpen
  \bibfield  {author} {\bibinfo {author} {\bibfnamefont {X.}~\bibnamefont
  {Li}}, \bibinfo {author} {\bibfnamefont {M.~C.}\ \bibnamefont {Kuzyk}}, \
  and\ \bibinfo {author} {\bibfnamefont {H.}~\bibnamefont {Wang}},\ }\href
  {\doibase 10.1103/PhysRevApplied.11.064037} {\bibfield  {journal} {\bibinfo
  {journal} {Phys. Rev. Applied}\ }\textbf {\bibinfo {volume} {11}},\ \bibinfo
  {pages} {064037} (\bibinfo {year} {2019})}\BibitemShut {NoStop}%
\bibitem [{\citenamefont {Maity}\ \emph {et~al.}(2019)\citenamefont {Maity},
  \citenamefont {Shao}, \citenamefont {Bogdanovi{\'c}}, \citenamefont
  {Meesala}, \citenamefont {Sohn}, \citenamefont {Sinclair}, \citenamefont
  {Pingault}, \citenamefont {Chalupnik}, \citenamefont {Chia}, \citenamefont
  {Zheng} \emph {et~al.}}]{maity2019coherent}%
  \BibitemOpen
  \bibfield  {author} {\bibinfo {author} {\bibfnamefont {S.}~\bibnamefont
  {Maity}}, \bibinfo {author} {\bibfnamefont {L.}~\bibnamefont {Shao}},
  \bibinfo {author} {\bibfnamefont {S.}~\bibnamefont {Bogdanovi{\'c}}},
  \bibinfo {author} {\bibfnamefont {S.}~\bibnamefont {Meesala}}, \bibinfo
  {author} {\bibfnamefont {Y.-I.}\ \bibnamefont {Sohn}}, \bibinfo {author}
  {\bibfnamefont {N.}~\bibnamefont {Sinclair}}, \bibinfo {author}
  {\bibfnamefont {B.}~\bibnamefont {Pingault}}, \bibinfo {author}
  {\bibfnamefont {M.}~\bibnamefont {Chalupnik}}, \bibinfo {author}
  {\bibfnamefont {C.}~\bibnamefont {Chia}}, \bibinfo {author} {\bibfnamefont
  {L.}~\bibnamefont {Zheng}},  \emph {et~al.},\ }\href@noop {} {\bibfield
  {journal} {\bibinfo  {journal} {arXiv preprint arXiv:1910.09710}\ } (\bibinfo
  {year} {2019})}\BibitemShut {NoStop}%
\bibitem [{\citenamefont {Lindner}\ \emph {et~al.}(2018)\citenamefont
  {Lindner}, \citenamefont {Bommer}, \citenamefont {Muzha}, \citenamefont
  {Krueger}, \citenamefont {Gines}, \citenamefont {Mandal}, \citenamefont
  {Williams}, \citenamefont {Londero}, \citenamefont {Gali},\ and\
  \citenamefont {Becher}}]{Lindner2018-ka}%
  \BibitemOpen
  \bibfield  {author} {\bibinfo {author} {\bibfnamefont {S.}~\bibnamefont
  {Lindner}}, \bibinfo {author} {\bibfnamefont {A.}~\bibnamefont {Bommer}},
  \bibinfo {author} {\bibfnamefont {A.}~\bibnamefont {Muzha}}, \bibinfo
  {author} {\bibfnamefont {A.}~\bibnamefont {Krueger}}, \bibinfo {author}
  {\bibfnamefont {L.}~\bibnamefont {Gines}}, \bibinfo {author} {\bibfnamefont
  {S.}~\bibnamefont {Mandal}}, \bibinfo {author} {\bibfnamefont
  {O.}~\bibnamefont {Williams}}, \bibinfo {author} {\bibfnamefont
  {E.}~\bibnamefont {Londero}}, \bibinfo {author} {\bibfnamefont
  {A.}~\bibnamefont {Gali}}, \ and\ \bibinfo {author} {\bibfnamefont
  {C.}~\bibnamefont {Becher}},\ }\href@noop {} {\bibfield  {journal} {\bibinfo
  {journal} {New J. Phys.}\ }\textbf {\bibinfo {volume} {20}},\ \bibinfo
  {pages} {115002} (\bibinfo {year} {2018})}\BibitemShut {NoStop}%
\bibitem [{\citenamefont {Simon}\ and\ \citenamefont
  {Irvine}(2003)}]{Simon2003-zu}%
  \BibitemOpen
  \bibfield  {author} {\bibinfo {author} {\bibfnamefont {C.}~\bibnamefont
  {Simon}}\ and\ \bibinfo {author} {\bibfnamefont {W.~T.~M.}\ \bibnamefont
  {Irvine}},\ }\href@noop {} {\bibfield  {journal} {\bibinfo  {journal} {Phys.
  Rev. Lett.}\ }\textbf {\bibinfo {volume} {91}},\ \bibinfo {pages} {110405}
  (\bibinfo {year} {2003})}\BibitemShut {NoStop}%
\bibitem [{\citenamefont {Moehring}\ \emph {et~al.}(2007)\citenamefont
  {Moehring}, \citenamefont {Maunz}, \citenamefont {Olmschenk}, \citenamefont
  {Younge}, \citenamefont {Matsukevich}, \citenamefont {Duan},\ and\
  \citenamefont {Monroe}}]{Moehring2007-qw}%
  \BibitemOpen
  \bibfield  {author} {\bibinfo {author} {\bibfnamefont {D.~L.}\ \bibnamefont
  {Moehring}}, \bibinfo {author} {\bibfnamefont {P.}~\bibnamefont {Maunz}},
  \bibinfo {author} {\bibfnamefont {S.}~\bibnamefont {Olmschenk}}, \bibinfo
  {author} {\bibfnamefont {K.~C.}\ \bibnamefont {Younge}}, \bibinfo {author}
  {\bibfnamefont {D.~N.}\ \bibnamefont {Matsukevich}}, \bibinfo {author}
  {\bibfnamefont {L.-M.}\ \bibnamefont {Duan}}, \ and\ \bibinfo {author}
  {\bibfnamefont {C.}~\bibnamefont {Monroe}},\ }\href@noop {} {\bibfield
  {journal} {\bibinfo  {journal} {Nature}\ }\textbf {\bibinfo {volume} {449}},\
  \bibinfo {pages} {68} (\bibinfo {year} {2007})}\BibitemShut {NoStop}%
\bibitem [{\citenamefont {Bernien}\ \emph {et~al.}(2013)\citenamefont
  {Bernien}, \citenamefont {Hensen}, \citenamefont {Pfaff}, \citenamefont
  {Koolstra}, \citenamefont {Blok}, \citenamefont {Robledo}, \citenamefont
  {Taminiau}, \citenamefont {Markham}, \citenamefont {Twitchen}, \citenamefont
  {Childress},\ and\ \citenamefont {Hanson}}]{Bernien2013-qt}%
  \BibitemOpen
  \bibfield  {author} {\bibinfo {author} {\bibfnamefont {H.}~\bibnamefont
  {Bernien}}, \bibinfo {author} {\bibfnamefont {B.}~\bibnamefont {Hensen}},
  \bibinfo {author} {\bibfnamefont {W.}~\bibnamefont {Pfaff}}, \bibinfo
  {author} {\bibfnamefont {G.}~\bibnamefont {Koolstra}}, \bibinfo {author}
  {\bibfnamefont {M.~S.}\ \bibnamefont {Blok}}, \bibinfo {author}
  {\bibfnamefont {L.}~\bibnamefont {Robledo}}, \bibinfo {author} {\bibfnamefont
  {T.~H.}\ \bibnamefont {Taminiau}}, \bibinfo {author} {\bibfnamefont
  {M.}~\bibnamefont {Markham}}, \bibinfo {author} {\bibfnamefont {D.~J.}\
  \bibnamefont {Twitchen}}, \bibinfo {author} {\bibfnamefont {L.}~\bibnamefont
  {Childress}}, \ and\ \bibinfo {author} {\bibfnamefont {R.}~\bibnamefont
  {Hanson}},\ }\href@noop {} {\bibfield  {journal} {\bibinfo  {journal}
  {Nature}\ }\textbf {\bibinfo {volume} {497}},\ \bibinfo {pages} {86}
  (\bibinfo {year} {2013})}\BibitemShut {NoStop}%
\bibitem [{\citenamefont {Dyckovsky}\ and\ \citenamefont
  {Olmschenk}(2012)}]{Dyckovsky2012-tf}%
  \BibitemOpen
  \bibfield  {author} {\bibinfo {author} {\bibfnamefont {A.~M.}\ \bibnamefont
  {Dyckovsky}}\ and\ \bibinfo {author} {\bibfnamefont {S.}~\bibnamefont
  {Olmschenk}},\ }\href@noop {} {\bibfield  {journal} {\bibinfo  {journal}
  {Phys. Rev. A}\ }\textbf {\bibinfo {volume} {85}},\ \bibinfo {pages} {052322}
  (\bibinfo {year} {2012})}\BibitemShut {NoStop}%
\bibitem [{\citenamefont {Lo~Piparo}\ \emph {et~al.}(2019)\citenamefont
  {Lo~Piparo}, \citenamefont {Munro},\ and\ \citenamefont
  {Nemoto}}]{Lo_Piparo2019-sa}%
  \BibitemOpen
  \bibfield  {author} {\bibinfo {author} {\bibfnamefont {N.}~\bibnamefont
  {Lo~Piparo}}, \bibinfo {author} {\bibfnamefont {W.~J.}\ \bibnamefont
  {Munro}}, \ and\ \bibinfo {author} {\bibfnamefont {K.}~\bibnamefont
  {Nemoto}},\ }\href@noop {} {\bibfield  {journal} {\bibinfo  {journal} {Phys.
  Rev. A}\ }\textbf {\bibinfo {volume} {99}},\ \bibinfo {pages} {022337}
  (\bibinfo {year} {2019})}\BibitemShut {NoStop}%
\bibitem [{\citenamefont {Atat{\"u}re}\ \emph {et~al.}(2018)\citenamefont
  {Atat{\"u}re}, \citenamefont {Englund}, \citenamefont {Vamivakas},
  \citenamefont {Lee},\ and\ \citenamefont {Wrachtrup}}]{Atature2018-oh}%
  \BibitemOpen
  \bibfield  {author} {\bibinfo {author} {\bibfnamefont {M.}~\bibnamefont
  {Atat{\"u}re}}, \bibinfo {author} {\bibfnamefont {D.}~\bibnamefont
  {Englund}}, \bibinfo {author} {\bibfnamefont {N.}~\bibnamefont {Vamivakas}},
  \bibinfo {author} {\bibfnamefont {S.-Y.}\ \bibnamefont {Lee}}, \ and\
  \bibinfo {author} {\bibfnamefont {J.}~\bibnamefont {Wrachtrup}},\ }\href@noop
  {} {\bibfield  {journal} {\bibinfo  {journal} {Nature Reviews Materials}\ }
  (\bibinfo {year} {2018})}\BibitemShut {NoStop}%
\bibitem [{\citenamefont {Nguyen}\ \emph {et~al.}(2019)\citenamefont {Nguyen},
  \citenamefont {Sukachev}, \citenamefont {Bhaskar}, \citenamefont {Machielse},
  \citenamefont {Levonian}, \citenamefont {Knall}, \citenamefont {Stroganov},
  \citenamefont {Chia}, \citenamefont {Burek}, \citenamefont {Riedinger},
  \citenamefont {Park}, \citenamefont {Lon{\v c}ar},\ and\ \citenamefont
  {Lukin}}]{Nguyen2019-yd}%
  \BibitemOpen
  \bibfield  {author} {\bibinfo {author} {\bibfnamefont {C.~T.}\ \bibnamefont
  {Nguyen}}, \bibinfo {author} {\bibfnamefont {D.~D.}\ \bibnamefont
  {Sukachev}}, \bibinfo {author} {\bibfnamefont {M.~K.}\ \bibnamefont
  {Bhaskar}}, \bibinfo {author} {\bibfnamefont {B.}~\bibnamefont {Machielse}},
  \bibinfo {author} {\bibfnamefont {D.~S.}\ \bibnamefont {Levonian}}, \bibinfo
  {author} {\bibfnamefont {E.~N.}\ \bibnamefont {Knall}}, \bibinfo {author}
  {\bibfnamefont {P.}~\bibnamefont {Stroganov}}, \bibinfo {author}
  {\bibfnamefont {C.}~\bibnamefont {Chia}}, \bibinfo {author} {\bibfnamefont
  {M.~J.}\ \bibnamefont {Burek}}, \bibinfo {author} {\bibfnamefont
  {R.}~\bibnamefont {Riedinger}}, \bibinfo {author} {\bibfnamefont
  {H.}~\bibnamefont {Park}}, \bibinfo {author} {\bibfnamefont {M.}~\bibnamefont
  {Lon{\v c}ar}}, \ and\ \bibinfo {author} {\bibfnamefont {M.~D.}\ \bibnamefont
  {Lukin}},\ }\href@noop {} {\bibfield  {journal} {\bibinfo  {journal} {Phys.
  Rev. B Condens. Matter}\ }\textbf {\bibinfo {volume} {100}},\ \bibinfo
  {pages} {165428} (\bibinfo {year} {2019})}\BibitemShut {NoStop}%
\bibitem [{\citenamefont {Meesala}\ \emph {et~al.}(2018)\citenamefont
  {Meesala}, \citenamefont {Sohn}, \citenamefont {Pingault}, \citenamefont
  {Shao}, \citenamefont {Atikian}, \citenamefont {Holzgrafe}, \citenamefont
  {G{\"u}ndo{\u g}an}, \citenamefont {Stavrakas}, \citenamefont {Sipahigil},
  \citenamefont {Chia}, \citenamefont {Evans}, \citenamefont {Burek},
  \citenamefont {Zhang}, \citenamefont {Wu}, \citenamefont {Pacheco},
  \citenamefont {Abraham}, \citenamefont {Bielejec}, \citenamefont {Lukin},
  \citenamefont {Atat{\"u}re},\ and\ \citenamefont {Lon{\v
  c}ar}}]{Meesala2018-zl}%
  \BibitemOpen
  \bibfield  {author} {\bibinfo {author} {\bibfnamefont {S.}~\bibnamefont
  {Meesala}}, \bibinfo {author} {\bibfnamefont {Y.-I.}\ \bibnamefont {Sohn}},
  \bibinfo {author} {\bibfnamefont {B.}~\bibnamefont {Pingault}}, \bibinfo
  {author} {\bibfnamefont {L.}~\bibnamefont {Shao}}, \bibinfo {author}
  {\bibfnamefont {H.~A.}\ \bibnamefont {Atikian}}, \bibinfo {author}
  {\bibfnamefont {J.}~\bibnamefont {Holzgrafe}}, \bibinfo {author}
  {\bibfnamefont {M.}~\bibnamefont {G{\"u}ndo{\u g}an}}, \bibinfo {author}
  {\bibfnamefont {C.}~\bibnamefont {Stavrakas}}, \bibinfo {author}
  {\bibfnamefont {A.}~\bibnamefont {Sipahigil}}, \bibinfo {author}
  {\bibfnamefont {C.}~\bibnamefont {Chia}}, \bibinfo {author} {\bibfnamefont
  {R.}~\bibnamefont {Evans}}, \bibinfo {author} {\bibfnamefont {M.~J.}\
  \bibnamefont {Burek}}, \bibinfo {author} {\bibfnamefont {M.}~\bibnamefont
  {Zhang}}, \bibinfo {author} {\bibfnamefont {L.}~\bibnamefont {Wu}}, \bibinfo
  {author} {\bibfnamefont {J.~L.}\ \bibnamefont {Pacheco}}, \bibinfo {author}
  {\bibfnamefont {J.}~\bibnamefont {Abraham}}, \bibinfo {author} {\bibfnamefont
  {E.}~\bibnamefont {Bielejec}}, \bibinfo {author} {\bibfnamefont {M.~D.}\
  \bibnamefont {Lukin}}, \bibinfo {author} {\bibfnamefont {M.}~\bibnamefont
  {Atat{\"u}re}}, \ and\ \bibinfo {author} {\bibfnamefont {M.}~\bibnamefont
  {Lon{\v c}ar}},\ }\href@noop {} {\bibfield  {journal} {\bibinfo  {journal}
  {Phys. Rev. B Condens. Matter}\ }\textbf {\bibinfo {volume} {97}},\ \bibinfo
  {pages} {205444} (\bibinfo {year} {2018})}\BibitemShut {NoStop}%
\bibitem [{\citenamefont {Noel}\ \emph {et~al.}(1998)\citenamefont {Noel},
  \citenamefont {Griffith},\ and\ \citenamefont {Gallagher}}]{noel1998freqmod}%
  \BibitemOpen
  \bibfield  {author} {\bibinfo {author} {\bibfnamefont {M.~W.}\ \bibnamefont
  {Noel}}, \bibinfo {author} {\bibfnamefont {W.~M.}\ \bibnamefont {Griffith}},
  \ and\ \bibinfo {author} {\bibfnamefont {T.~F.}\ \bibnamefont {Gallagher}},\
  }\href {\doibase 10.1103/PhysRevA.58.2265} {\bibfield  {journal} {\bibinfo
  {journal} {Phys. Rev. A}\ }\textbf {\bibinfo {volume} {58}},\ \bibinfo
  {pages} {2265} (\bibinfo {year} {1998})}\BibitemShut {NoStop}%
\bibitem [{\citenamefont {Ashhab}\ \emph {et~al.}(2007)\citenamefont {Ashhab},
  \citenamefont {Johansson}, \citenamefont {Zagoskin},\ and\ \citenamefont
  {Nori}}]{ashhab2007drivetheory}%
  \BibitemOpen
  \bibfield  {author} {\bibinfo {author} {\bibfnamefont {S.}~\bibnamefont
  {Ashhab}}, \bibinfo {author} {\bibfnamefont {J.~R.}\ \bibnamefont
  {Johansson}}, \bibinfo {author} {\bibfnamefont {A.~M.}\ \bibnamefont
  {Zagoskin}}, \ and\ \bibinfo {author} {\bibfnamefont {F.}~\bibnamefont
  {Nori}},\ }\href {\doibase 10.1103/PhysRevA.75.063414} {\bibfield  {journal}
  {\bibinfo  {journal} {Phys. Rev. A}\ }\textbf {\bibinfo {volume} {75}},\
  \bibinfo {pages} {063414} (\bibinfo {year} {2007})}\BibitemShut {NoStop}%
\bibitem [{\citenamefont {Beaudoin}\ \emph {et~al.}(2012)\citenamefont
  {Beaudoin}, \citenamefont {da~Silva}, \citenamefont {Dutton},\ and\
  \citenamefont {Blais}}]{beaudoin2012sidebands}%
  \BibitemOpen
  \bibfield  {author} {\bibinfo {author} {\bibfnamefont {F.}~\bibnamefont
  {Beaudoin}}, \bibinfo {author} {\bibfnamefont {M.~P.}\ \bibnamefont
  {da~Silva}}, \bibinfo {author} {\bibfnamefont {Z.}~\bibnamefont {Dutton}}, \
  and\ \bibinfo {author} {\bibfnamefont {A.}~\bibnamefont {Blais}},\ }\href
  {\doibase 10.1103/PhysRevA.86.022305} {\bibfield  {journal} {\bibinfo
  {journal} {Phys. Rev. A}\ }\textbf {\bibinfo {volume} {86}},\ \bibinfo
  {pages} {022305} (\bibinfo {year} {2012})}\BibitemShut {NoStop}%
\bibitem [{\citenamefont {Strand}\ \emph {et~al.}(2013)\citenamefont {Strand},
  \citenamefont {Ware}, \citenamefont {Beaudoin}, \citenamefont {Ohki},
  \citenamefont {Johnson}, \citenamefont {Blais},\ and\ \citenamefont
  {Plourde}}]{strand2013sideband}%
  \BibitemOpen
  \bibfield  {author} {\bibinfo {author} {\bibfnamefont {J.~D.}\ \bibnamefont
  {Strand}}, \bibinfo {author} {\bibfnamefont {M.}~\bibnamefont {Ware}},
  \bibinfo {author} {\bibfnamefont {F.}~\bibnamefont {Beaudoin}}, \bibinfo
  {author} {\bibfnamefont {T.~A.}\ \bibnamefont {Ohki}}, \bibinfo {author}
  {\bibfnamefont {B.~R.}\ \bibnamefont {Johnson}}, \bibinfo {author}
  {\bibfnamefont {A.}~\bibnamefont {Blais}}, \ and\ \bibinfo {author}
  {\bibfnamefont {B.~L.~T.}\ \bibnamefont {Plourde}},\ }\href {\doibase
  10.1103/PhysRevB.87.220505} {\bibfield  {journal} {\bibinfo  {journal} {Phys.
  Rev. B}\ }\textbf {\bibinfo {volume} {87}},\ \bibinfo {pages} {220505}
  (\bibinfo {year} {2013})}\BibitemShut {NoStop}%
\bibitem [{\citenamefont {Silveri}\ \emph {et~al.}(2017)\citenamefont
  {Silveri}, \citenamefont {Tuorila}, \citenamefont {Thuneberg},\ and\
  \citenamefont {Paraoanu}}]{silveri2017quantum}%
  \BibitemOpen
  \bibfield  {author} {\bibinfo {author} {\bibfnamefont {M.}~\bibnamefont
  {Silveri}}, \bibinfo {author} {\bibfnamefont {J.}~\bibnamefont {Tuorila}},
  \bibinfo {author} {\bibfnamefont {E.}~\bibnamefont {Thuneberg}}, \ and\
  \bibinfo {author} {\bibfnamefont {G.}~\bibnamefont {Paraoanu}},\ }\href@noop
  {} {\bibfield  {journal} {\bibinfo  {journal} {Rep. Prog. Phys.}\ }\textbf
  {\bibinfo {volume} {80}},\ \bibinfo {pages} {056002} (\bibinfo {year}
  {2017})}\BibitemShut {NoStop}%
\bibitem [{\citenamefont {Oliver}\ \emph {et~al.}(2005)\citenamefont {Oliver},
  \citenamefont {Yu}, \citenamefont {Lee}, \citenamefont {Berggren},
  \citenamefont {Levitov},\ and\ \citenamefont
  {Orlando}}]{Oliver1653machzender}%
  \BibitemOpen
  \bibfield  {author} {\bibinfo {author} {\bibfnamefont {W.~D.}\ \bibnamefont
  {Oliver}}, \bibinfo {author} {\bibfnamefont {Y.}~\bibnamefont {Yu}}, \bibinfo
  {author} {\bibfnamefont {J.~C.}\ \bibnamefont {Lee}}, \bibinfo {author}
  {\bibfnamefont {K.~K.}\ \bibnamefont {Berggren}}, \bibinfo {author}
  {\bibfnamefont {L.~S.}\ \bibnamefont {Levitov}}, \ and\ \bibinfo {author}
  {\bibfnamefont {T.~P.}\ \bibnamefont {Orlando}},\ }\href {\doibase
  10.1126/science.1119678} {\bibfield  {journal} {\bibinfo  {journal}
  {Science}\ }\textbf {\bibinfo {volume} {310}},\ \bibinfo {pages} {1653}
  (\bibinfo {year} {2005})}\BibitemShut {NoStop}%
\bibitem [{\citenamefont {Maity}\ \emph {et~al.}(2018)\citenamefont {Maity},
  \citenamefont {Shao}, \citenamefont {Sohn}, \citenamefont {Meesala},
  \citenamefont {Machielse}, \citenamefont {Bielejec}, \citenamefont
  {Markham},\ and\ \citenamefont {Lon\ifmmode~\check{c}\else
  \v{c}\fi{}ar}}]{maity2018alignment}%
  \BibitemOpen
  \bibfield  {author} {\bibinfo {author} {\bibfnamefont {S.}~\bibnamefont
  {Maity}}, \bibinfo {author} {\bibfnamefont {L.}~\bibnamefont {Shao}},
  \bibinfo {author} {\bibfnamefont {Y.-I.}\ \bibnamefont {Sohn}}, \bibinfo
  {author} {\bibfnamefont {S.}~\bibnamefont {Meesala}}, \bibinfo {author}
  {\bibfnamefont {B.}~\bibnamefont {Machielse}}, \bibinfo {author}
  {\bibfnamefont {E.}~\bibnamefont {Bielejec}}, \bibinfo {author}
  {\bibfnamefont {M.}~\bibnamefont {Markham}}, \ and\ \bibinfo {author}
  {\bibfnamefont {M.}~\bibnamefont {Lon\ifmmode~\check{c}\else \v{c}\fi{}ar}},\
  }\href {\doibase 10.1103/PhysRevApplied.10.024050} {\bibfield  {journal}
  {\bibinfo  {journal} {Phys. Rev. Applied}\ }\textbf {\bibinfo {volume}
  {10}},\ \bibinfo {pages} {024050} (\bibinfo {year} {2018})}\BibitemShut
  {NoStop}%
\bibitem [{\citenamefont {Udvarhelyi}\ \emph {et~al.}(2018)\citenamefont
  {Udvarhelyi}, \citenamefont {Shkolnikov}, \citenamefont {Gali}, \citenamefont
  {Burkard},\ and\ \citenamefont {P\'alyi}}]{udvarheliy2018spinstrain}%
  \BibitemOpen
  \bibfield  {author} {\bibinfo {author} {\bibfnamefont {P.}~\bibnamefont
  {Udvarhelyi}}, \bibinfo {author} {\bibfnamefont {V.~O.}\ \bibnamefont
  {Shkolnikov}}, \bibinfo {author} {\bibfnamefont {A.}~\bibnamefont {Gali}},
  \bibinfo {author} {\bibfnamefont {G.}~\bibnamefont {Burkard}}, \ and\
  \bibinfo {author} {\bibfnamefont {A.}~\bibnamefont {P\'alyi}},\ }\href
  {\doibase 10.1103/PhysRevB.98.075201} {\bibfield  {journal} {\bibinfo
  {journal} {Phys. Rev. B}\ }\textbf {\bibinfo {volume} {98}},\ \bibinfo
  {pages} {075201} (\bibinfo {year} {2018})}\BibitemShut {NoStop}%
\bibitem [{\citenamefont {Chen}\ \emph {et~al.}(2018)\citenamefont {Chen},
  \citenamefont {MacQuarrie},\ and\ \citenamefont {Fuchs}}]{chen2018orbital}%
  \BibitemOpen
  \bibfield  {author} {\bibinfo {author} {\bibfnamefont {H.~Y.}\ \bibnamefont
  {Chen}}, \bibinfo {author} {\bibfnamefont {E.~R.}\ \bibnamefont
  {MacQuarrie}}, \ and\ \bibinfo {author} {\bibfnamefont {G.~D.}\ \bibnamefont
  {Fuchs}},\ }\href {\doibase 10.1103/PhysRevLett.120.167401} {\bibfield
  {journal} {\bibinfo  {journal} {Phys. Rev. Lett.}\ }\textbf {\bibinfo
  {volume} {120}},\ \bibinfo {pages} {167401} (\bibinfo {year}
  {2018})}\BibitemShut {NoStop}%
\bibitem [{\citenamefont {Dmitriev}\ \emph {et~al.}(2019)\citenamefont
  {Dmitriev}, \citenamefont {Chen}, \citenamefont {Fuchs},\ and\ \citenamefont
  {Vershovskii}}]{dmitriev2019spinresonance}%
  \BibitemOpen
  \bibfield  {author} {\bibinfo {author} {\bibfnamefont {A.~K.}\ \bibnamefont
  {Dmitriev}}, \bibinfo {author} {\bibfnamefont {H.~Y.}\ \bibnamefont {Chen}},
  \bibinfo {author} {\bibfnamefont {G.~D.}\ \bibnamefont {Fuchs}}, \ and\
  \bibinfo {author} {\bibfnamefont {A.~K.}\ \bibnamefont {Vershovskii}},\
  }\href {\doibase 10.1103/PhysRevA.100.011801} {\bibfield  {journal} {\bibinfo
   {journal} {Phys. Rev. A}\ }\textbf {\bibinfo {volume} {100}},\ \bibinfo
  {pages} {011801} (\bibinfo {year} {2019})}\BibitemShut {NoStop}%
\bibitem [{\citenamefont {Breuer}\ and\ \citenamefont
  {Petruccione}(2003)}]{Breuer2003}%
  \BibitemOpen
  \bibfield  {author} {\bibinfo {author} {\bibfnamefont {H.-P.}\ \bibnamefont
  {Breuer}}\ and\ \bibinfo {author} {\bibfnamefont {F.}~\bibnamefont
  {Petruccione}},\ }\href@noop {} {\emph {\bibinfo {title} {The theory of open
  quantum systems}}}\ (\bibinfo  {publisher} {Oxford {U}niversity {P}ress},\
  \bibinfo {year} {2003})\BibitemShut {NoStop}%
\bibitem [{\citenamefont {Zheng}\ and\ \citenamefont
  {Guo}(2000)}]{zheng2000twoatentangl}%
  \BibitemOpen
  \bibfield  {author} {\bibinfo {author} {\bibfnamefont {S.-B.}\ \bibnamefont
  {Zheng}}\ and\ \bibinfo {author} {\bibfnamefont {G.-C.}\ \bibnamefont
  {Guo}},\ }\href {\doibase 10.1103/PhysRevLett.85.2392} {\bibfield  {journal}
  {\bibinfo  {journal} {Phys. Rev. Lett.}\ }\textbf {\bibinfo {volume} {85}},\
  \bibinfo {pages} {2392} (\bibinfo {year} {2000})}\BibitemShut {NoStop}%
\bibitem [{\citenamefont {Blais}\ \emph {et~al.}(2004)\citenamefont {Blais},
  \citenamefont {Huang}, \citenamefont {Wallraff}, \citenamefont {Girvin},\
  and\ \citenamefont {Schoelkopf}}]{blais2004dispersive}%
  \BibitemOpen
  \bibfield  {author} {\bibinfo {author} {\bibfnamefont {A.}~\bibnamefont
  {Blais}}, \bibinfo {author} {\bibfnamefont {R.-S.}\ \bibnamefont {Huang}},
  \bibinfo {author} {\bibfnamefont {A.}~\bibnamefont {Wallraff}}, \bibinfo
  {author} {\bibfnamefont {S.~M.}\ \bibnamefont {Girvin}}, \ and\ \bibinfo
  {author} {\bibfnamefont {R.~J.}\ \bibnamefont {Schoelkopf}},\ }\href
  {\doibase 10.1103/PhysRevA.69.062320} {\bibfield  {journal} {\bibinfo
  {journal} {Phys. Rev. A}\ }\textbf {\bibinfo {volume} {69}},\ \bibinfo
  {pages} {062320} (\bibinfo {year} {2004})}\BibitemShut {NoStop}%
\bibitem [{\citenamefont {Majer}\ \emph {et~al.}(2007)\citenamefont {Majer},
  \citenamefont {Chow}, \citenamefont {Gambetta}, \citenamefont {Koch},
  \citenamefont {Johnson}, \citenamefont {Schreier}, \citenamefont {Frunzio},
  \citenamefont {Schuster}, \citenamefont {Houck}, \citenamefont {Wallraff},
  \citenamefont {Blais}, \citenamefont {Devoret}, \citenamefont {Girvin},\ and\
  \citenamefont {Schoelkopf}}]{majer2007cavitybus}%
  \BibitemOpen
  \bibfield  {author} {\bibinfo {author} {\bibfnamefont {J.}~\bibnamefont
  {Majer}}, \bibinfo {author} {\bibfnamefont {J.}~\bibnamefont {Chow}},
  \bibinfo {author} {\bibfnamefont {J.}~\bibnamefont {Gambetta}}, \bibinfo
  {author} {\bibfnamefont {J.}~\bibnamefont {Koch}}, \bibinfo {author}
  {\bibfnamefont {B.}~\bibnamefont {Johnson}}, \bibinfo {author} {\bibfnamefont
  {J.}~\bibnamefont {Schreier}}, \bibinfo {author} {\bibfnamefont
  {L.}~\bibnamefont {Frunzio}}, \bibinfo {author} {\bibfnamefont
  {D.}~\bibnamefont {Schuster}}, \bibinfo {author} {\bibfnamefont
  {A.}~\bibnamefont {Houck}}, \bibinfo {author} {\bibfnamefont
  {A.}~\bibnamefont {Wallraff}}, \bibinfo {author} {\bibfnamefont
  {A.}~\bibnamefont {Blais}}, \bibinfo {author} {\bibfnamefont
  {M.}~\bibnamefont {Devoret}}, \bibinfo {author} {\bibfnamefont
  {S.}~\bibnamefont {Girvin}}, \ and\ \bibinfo {author} {\bibfnamefont
  {R.}~\bibnamefont {Schoelkopf}},\ }\href {\doibase 10.1038/nature06184}
  {\bibfield  {journal} {\bibinfo  {journal} {Nature}\ }\textbf {\bibinfo
  {volume} {449}},\ \bibinfo {pages} {443} (\bibinfo {year}
  {2007})}\BibitemShut {NoStop}%
\bibitem [{\citenamefont {Abramowitz}\ and\ \citenamefont
  {Stegun}(1964)}]{abramowitz65functions}%
  \BibitemOpen
  \bibfield  {author} {\bibinfo {author} {\bibfnamefont {M.}~\bibnamefont
  {Abramowitz}}\ and\ \bibinfo {author} {\bibfnamefont {I.~A.}\ \bibnamefont
  {Stegun}},\ }\href@noop {} {\emph {\bibinfo {title} {{H}andbook of
  Mathematical Functions with Formulas, Graphs, and Mathematical Tables}}}\
  (\bibinfo  {publisher} {National Bureau of Standards Applied Mathematics
  Series 55},\ \bibinfo {year} {1964})\BibitemShut {NoStop}%
\bibitem [{\citenamefont {Anderson}\ \emph {et~al.}(2019)\citenamefont
  {Anderson}, \citenamefont {Bourassa}, \citenamefont {Miao}, \citenamefont
  {Wolfowicz}, \citenamefont {Mintun}, \citenamefont {Crook}, \citenamefont
  {Abe}, \citenamefont {Ul~Hassan}, \citenamefont {Son}, \citenamefont
  {Ohshima},\ and\ \citenamefont {Awschalom}}]{Anderson2019-wo}%
  \BibitemOpen
  \bibfield  {author} {\bibinfo {author} {\bibfnamefont {C.~P.}\ \bibnamefont
  {Anderson}}, \bibinfo {author} {\bibfnamefont {A.}~\bibnamefont {Bourassa}},
  \bibinfo {author} {\bibfnamefont {K.~C.}\ \bibnamefont {Miao}}, \bibinfo
  {author} {\bibfnamefont {G.}~\bibnamefont {Wolfowicz}}, \bibinfo {author}
  {\bibfnamefont {P.~J.}\ \bibnamefont {Mintun}}, \bibinfo {author}
  {\bibfnamefont {A.~L.}\ \bibnamefont {Crook}}, \bibinfo {author}
  {\bibfnamefont {H.}~\bibnamefont {Abe}}, \bibinfo {author} {\bibfnamefont
  {J.}~\bibnamefont {Ul~Hassan}}, \bibinfo {author} {\bibfnamefont {N.~T.}\
  \bibnamefont {Son}}, \bibinfo {author} {\bibfnamefont {T.}~\bibnamefont
  {Ohshima}}, \ and\ \bibinfo {author} {\bibfnamefont {D.~D.}\ \bibnamefont
  {Awschalom}},\ }\href {\doibase 10.1126/science.aax9406} {\bibfield
  {journal} {\bibinfo  {journal} {Science}\ }\textbf {\bibinfo {volume}
  {366}},\ \bibinfo {pages} {1225} (\bibinfo {year} {2019})}\BibitemShut
  {NoStop}%
\bibitem [{\citenamefont {Asadi}\ \emph {et~al.}(2019)\citenamefont {Asadi},
  \citenamefont {Wein},\ and\ \citenamefont {Simon}}]{Asadi2019-lc}%
  \BibitemOpen
  \bibfield  {author} {\bibinfo {author} {\bibfnamefont {F.~K.}\ \bibnamefont
  {Asadi}}, \bibinfo {author} {\bibfnamefont {S.}~\bibnamefont {Wein}}, \ and\
  \bibinfo {author} {\bibfnamefont {C.}~\bibnamefont {Simon}},\ }\href@noop {}
  {\  (\bibinfo {year} {2019})},\ \Eprint {http://arxiv.org/abs/1911.02176}
  {arXiv:1911.02176 [quant-ph]} \BibitemShut {NoStop}%
\bibitem [{\citenamefont {MacQuarrie}\ \emph {et~al.}(2013)\citenamefont
  {MacQuarrie}, \citenamefont {Gosavi}, \citenamefont {Jungwirth},
  \citenamefont {Bhave},\ and\ \citenamefont
  {Fuchs}}]{MacQuarrie2013mechanical}%
  \BibitemOpen
  \bibfield  {author} {\bibinfo {author} {\bibfnamefont {E.~R.}\ \bibnamefont
  {MacQuarrie}}, \bibinfo {author} {\bibfnamefont {T.~A.}\ \bibnamefont
  {Gosavi}}, \bibinfo {author} {\bibfnamefont {N.~R.}\ \bibnamefont
  {Jungwirth}}, \bibinfo {author} {\bibfnamefont {S.~A.}\ \bibnamefont
  {Bhave}}, \ and\ \bibinfo {author} {\bibfnamefont {G.~D.}\ \bibnamefont
  {Fuchs}},\ }\href {\doibase 10.1103/PhysRevLett.111.227602} {\bibfield
  {journal} {\bibinfo  {journal} {Phys. Rev. Lett.}\ }\textbf {\bibinfo
  {volume} {111}},\ \bibinfo {pages} {227602} (\bibinfo {year}
  {2013})}\BibitemShut {NoStop}%
\end{thebibliography}%


\newcommand{\noopsort}[1]{} \newcommand{\printfirst}[2]{#1}
  \newcommand{\singleletter}[1]{#1} \newcommand{\switchargs}[2]{#2#1}
%
\end{document}